\documentclass[acmsmall,screen]{acmart}
\usepackage{tabularx}
\usepackage{multirow} 
\usepackage{threeparttable} 
\usepackage{wasysym}
\usepackage{color,soul}
\usepackage{xcolor}
\usepackage{listings}
\usepackage{subcaption}
\usepackage{multirow}
\usepackage{changepage}
\usepackage{threeparttable}
\usepackage{booktabs}
\usepackage{tcolorbox} 
\usepackage{fontawesome5} 
\usepackage{colortbl}
\usepackage{tabularx}
\usepackage{colortbl}

\newtcolorbox{boxK}{
    top=2pt,
    bottom=2pt,
    left=2pt,
    right=2pt,
    boxrule = 0pt,
    toprule = 0pt, 
}

\definecolor{codeblue}{RGB}{86, 168, 245}   
\definecolor{codebrown}{RGB}{170, 73, 38}   
\definecolor{codepurple}{RGB}{136,136,198} 
\definecolor{codegreen}{RGB}{106,171,115}    
\definecolor{codeorange}{RGB}{207, 142, 109}  

\definecolor{instr}{RGB}{0, 80, 160}    
\definecolor{code}{RGB}{180, 0, 40}     
\definecolor{output}{RGB}{0, 120, 60}    
\definecolor{example}{RGB}{100, 60, 160} 

\newtcolorbox{promptbox}[1][]{colback=gray!5!white,
  colframe=black!80,
  boxrule=0.6pt,
  arc=2mm,
  left=2mm,
  right=2mm,
  top=1mm,
  bottom=1mm,
  fontupper=\ttfamily\footnotesize ,
  #1}
\AtBeginDocument{%
  }

\setcopyright{acmlicensed}
\copyrightyear{2018}
\acmYear{2018}
\acmDOI{XXXXXXX.XXXXXXX}
\acmConference[Conference acronym 'XX]{Make sure to enter the correct
  conference title from your rights confirmation email}{June 03--05,
  2018}{Woodstock, NY}
\acmISBN{978-1-4503-XXXX-X/2018/06}

\soulregister{\cite}7 
\soulregister{\citep}7 
\soulregister{\citet}7 
\soulregister{\ref}7 
\soulregister{\pageref}7 
\begin{document}

\title{\textit{TransLibEval}: Demystify Large Language Models' Capability in Third-party Library-targeted Code Translation}

\author{Pengyu Xue}
\email{xuepengyu@mail.sdu.edu.cn}
\orcid{0009-0007-3395-9575}
\authornote{These authors contributed equally to this work.}
\affiliation{%
   \institution{Shandong University}
   \city{Qingdao}
   \state{Shandong}
   \country{China}
}

\author{Kunwu Zheng}
\email{Xiaozhengsdu2022@mail.sdu.edu.cn}
\orcid{0009-0000-7047-1981}
\authornotemark[1]
\affiliation{%
   \institution{Shandong University}
   \city{Qingdao}
   \state{Shandong}
   \country{China}
}

\author{Zhen Yang}
\email{zhenyang@sdu.edu.cn}
\orcid{0000-0003-0670-4538}
\authornote{Corresponding author.}
\affiliation{%
   \institution{Shandong University}
   \city{Qingdao}
   \state{Shandong}
   \country{China}
}

\author{Yifei Pei}
\email{peiyifei@mail.sdu.edu.cn}
\orcid{0009-0005-1351-0565}
\affiliation{%
   \institution{Shandong University}
   \city{Qingdao}
   \state{Shandong}
   \country{China}
}

\author{Linhao Wu}
\email{wulinhao@mail.sdu.edu.cn}
\orcid{0009-0001-7624-156X}
\affiliation{%
   \institution{Shandong University}
   \city{Qingdao}
   \state{Shandong}
   \country{China}
}

\author{Jiahui Dong}
\email{202400130128@mail.sdu.edu.cn}
\orcid{0009-0002-1307-3981}
\affiliation{%
   \institution{Shandong University}
   \city{Qingdao}
   \state{Shandong}
   \country{China}
}

\author{Xiapu Luo}
\email{csxluo@comp.polyu.edu.hk}
\orcid{0000-0002-9082-3208}
\affiliation{%
   \institution{The Hong Kong Polytechnic University}
   \state{Hong Kong}
   \country{China}
}

\author{Yan Xiao}
\email{xiaoyan.hhu@gmail.com}
\orcid{0000-0002-2563-083X}
\affiliation{%
   \institution{Sun Yat-sen University}
   \city{Shenzhen}
   \state{Guangdong}
   \country{China}
}

\author{Fei Liu}
\email{15066345378@163.com}
\orcid{0009-0002-5722-1990}
\affiliation{%
   \institution{Shandong University}
   \city{Qingdao}
   \state{Shandong}
   \country{China}
}

\author{Yuxuan Zhang}
\email{202400130071@mail.sdu.edu.cn}
\orcid{0009-0007-4791-3770}
\affiliation{%
   \institution{Shandong University}
   \city{Qingdao}
   \state{Shandong}
   \country{China}
}

\author{Xiran Lyu}
\email{202400130069@mail.sdu.edu.cn}
\orcid{0009-0002-8861-8907}
\affiliation{%
   \institution{Shandong University}
   \city{Qingdao}
   \state{Shandong}
   \country{China}
}

\author{Xianhang Li}
\email{202400130007@mail.sdu.edu.cn}
\orcid{0009-0006-8852-1215}
\affiliation{%
   \institution{Shandong University}
   \city{Qingdao}
   \state{Shandong}
   \country{China}
}

\author{Xuanyu Zhu}
\email{zhuxuanyu@mail.sdu.edu.cn}
\orcid{0009-0005-9896-1647}
\affiliation{%
   \institution{Shandong University}
   \city{Qingdao}
   \state{Shandong}
   \country{China}
}

\author{Chengyi Wang}
\email{202300130150@mail.sdu.edu.cn}
\orcid{0009-0000-4153-6774}
\affiliation{%
   \institution{Shandong University}
   \city{Qingdao}
   \state{Shandong}
   \country{China}
}
\renewcommand{\shortauthors}{Xue et al.}

\begin{abstract}

  In recent years, Large Language Models (LLMs) have been widely studied in the code translation field on the method, class, and even repository levels. 
  However, most of these benchmarks are limited in terms of Third-Party Library (TPL) categories and scales, making TPL-related errors hard to expose and hindering the development of targeted solutions. Considering the high dependence (over 90\%) on TPLs in practical programming, demystifying and analyzing LLMs' code translation performance involving various TPLs becomes imperative.    

  To address this gap, we construct \textit{TransLibEval}, the first benchmark dedicated to library-centric code translation. It consists of 200 real-world tasks across Python, Java, and C++, each explicitly involving TPLs from diverse categories such as data processing, machine learning, and web development, with comprehensive dependency coverage and high-coverage test suites. We evaluate seven recent LLMs of commercial, general, and code-specialized families under six translation strategies of three categories: Direct, IR-guided, and Retrieval-augmented. Experimental results show a dramatic performance drop compared with library-free settings (average CA decline over 60\%), while diverse strategies demonstrate heterogeneous advantages. 
  Furthermore, we analyze 4,831 failed cases from GPT-4o, one of the State-of-the-Art (SOTA) LLMs, revealing numerous third-party reference errors that were obscured previously. These findings highlight the unique challenges of library-centric translation and provide practical guidance for improving TPL-aware code intelligence.
  
\end{abstract}



\maketitle
\vspace{-0.2cm}
\section{Introduction}

{Automated code translation} has become increasingly important for modern software engineering, as it enables developers to port codebases across multiple Programming Languages (PLs) to adapt to diverse platforms, ranging from cloud services and web applications to embedded systems. In recent years, LLMs, such as GPT-4o and DeepSeek-V3, have been widely studied and applied in code translation tasks across diverse contexts from standalone functions \cite{roziere2023code, tao2024unraveling, yang2024exploring, yuan2024transagent} and class-level designs \cite{xue2024classeval-t} to repository-level environments \cite{ou2024repository, ibrahimzada2024repository}. 


However, most existing benchmarks remain limited in their coverage of both the variety and scale of Third-Party Libraries (TPLs), owing to the reasons we analyzed in Section \ref{Related Work}, which significantly restricts the exposure of translation errors rooted in external TPLs. 
Nonetheless, in real-world programming, over 90\% of development scenarios necessitate the involvement of TPLs \cite{huang2022characterizing}, including tasks of data processing, machine learning, web application development, and database interaction.
Translating code across languages in these scenarios introduces additional challenges, such as recognizing the correct TPLs, adapting cross-language library mappings, and handling library-specific APIs/parameter conventions. Current benchmarks without limited TPLs cannot adequately reflect these practical challenges. This motivates the creation of a new benchmark that directly measures LLMs’ ability to handle external TPLs in code translation tasks.

\textbf{Benchmark \textit{TransLibEval}:} To mitigate the above limitations, we propose \textit{TransLibEval}, the first benchmark specifically designed for evaluating code translation with TPLs. \textit{TransLibEval} consists of 200 real-world translation tasks across Python, Java, and C++, where every task explicitly involves TPLs, covering ten categories such as data processing, utilities, web development, machine learning, and security. Each task is constructed from Python as the calibration PL, then manually translated into Java and C++ by experienced software engineers following strict style and type conventions. \textit{TransLibEval} features 
(1) more lines of code than most parallel benchmarks with an 88\% lift on average.
(2) With 5 tests per task, \textit{TransLibEval} achieves 95\% statement coverage, significantly improving evaluation efficiency;
(3) Existing benchmarks exhibit extremely low library dependency rates (mostly <11\%), whereas \textit{TransLibEval} covers 200 libraries across 10 categories, effectively filling a research gap.

\textbf{Empirical Study:} Based on \textit{TransLibEval}, we conduct the first large-scale evaluation of recent LLMs on library-centric code translation. {Specifically, we assess seven representative LLMs, including commercial (GPT-4o, GPT-3.5, DeepSeek-V3, Qwen-Max), general-purpose (Llama3-8B/70B), and code-specialized (CodeLlama) models.} To capture translation behaviors, we investigate six translation strategies of three families: (1) \emph{Direct translation}, where LLMs translate code snippets with TPLs end-to-end; (2) \emph{IR-guided translation}, which generates intermediate representations (e.g., pseudocode, reasoning traces, or summaries) to guide the final output; and (3) \emph{Retrieval-augmented translation}, where external knowledge from Stack Overflow \cite{stackoverflow} posts is retrieved to aid library selection and API usage. We evaluate each program with Compilation Success Rate (CSR), Computational Accuracy (CA),  Pass Rate (PR), and Library Dependency Awareness (LDA) to measure whether correct third-party libraries are identified and used.

\textbf{Main findings:} Our experiments reveal several insights. (1) {LLMs perform significantly worse on library-centric code translation than on library-free benchmarks, with CA dropping by over 60\% on average, highlighting the difficulty of handling TPLs.} (2) {Commercial LLMs outperform smaller LLMs, with GPT-4o and DeepSeek leading across most metrics, while smaller models like CodeLlama show limited performance despite code specialization.} (3) {Translations into Python are consistently easier for LLMs, achieving over 50\% CA, while translations into Java and C++ remain highly challenging due to stricter type systems and fewer available TPLs.} (4) {Strategy selection matters: \emph{IR(pseudocode)} performs best when translating into Python, \emph{RA(name)} is more effective for C++, while \emph{RA(method)} and \emph{Direct} excel in Java translations.} (5) {Error analysis uncovers a great amount of TPL reference errors, where API-related errors, especially calling an irrelevant API, account for over 50\% of the total share, while different errors also skewed to diverse translation strategies (see details in Section \ref{RQ4: Failed Cases Analysis}).
The contributions of this work are threefold:

(1) We construct the first library-centric code translation benchmark, \textit{TransLibEval}, covering 200 real-world TPL-based tasks of 10 categories across Python, Java, and C++. The benchmark dataset and associated code are available at \cite{TransLib77:online}.

(2) We provide a systematic evaluation of diverse LLMs and translation strategies, combining correctness, compilation, and dependency-awareness metrics.

(3) We perform a large-scale error categorization of 4,831 failed cases, supported by our detailed 21-page Failed Cases Report \cite{TransLibEvalError2026}. Besides, failures of each translation strategy are also revealed, offering new insights into the challenges of TPL translation and directions for future improvements.

\section{Related Work}
\label{Related Work}

\subsection{Large Language Models for Code Translation}

In recent years, substantial effort has been dedicated to investigating and enhancing LLM-based code translation. Initial studies focused on analyzing LLM performance and error types~\cite{Pan2023Effectiveness, Eniser2023Rust}, while others explored methods to improve translation fidelity, such as leveraging test cases (UniTrans~\cite{yang2024exploring}), employing multi-agent fixing (TRANSAGENT~\cite{yuan2024transagent1}), or developing specialized models (SteloCoder~\cite{pan2023stelocoder}), integrated API knowledge \cite{Wang2025ApiRAT}, and general correctors (Rectifier~\cite{yinxin2024rectifier}) for algorithmic function-level translation.  

Beyond standalone functions, recent work has begun to explore code translation at the repository level, addressing real-world complexities like inter-file dependencies and larger codebases \cite{ibrahimzada2024repository,wang2025evoc2rust,Ou2025KTrans,yuan2024transagent1,zhou2025porting,gao2024rule}. This includes efforts to augment LLMs with extra knowledge, e.g., K$^3$Trans~\cite{Ou2025KTrans}, and with multi-agent systems, e.g., RepoTransAgent~\cite{yuan2024transagent1}.
However, owing to the high complexity of code repositories, the studies above are either constrained in repository amount ($\leq$6) \cite{wang2025evoc2rust,yuan2024transagent1} or even proactively exclude TPLs to reduce translation difficulties \cite{ibrahimzada2024repository,yuan2024transagent1}. Certain studies focus on domain-specific PLs (e.g., Rust or Swift), restricting the spectrum of library categories to limited areas \cite{Ou2025KTrans,zhou2025porting,gao2024rule}. 
Thus, their empirical evaluations often do not specifically focus on a broad and diverse range of TPL integrations. 
This leaves a notable gap in understanding LLM capabilities in TPL-targeted code translation scenarios, which account for over 90\% in real-world development \cite{huang2022characterizing}. As such, this work introduces \textit{TransLibEval}, a benchmark specifically engineered to assess code translation performance on library-centric tasks.


\subsection{Existing Benchmarks for Code Translation}
Code translation techniques constitute an efficient process for cross-language codebase migration, which is inseparable from a complete evaluation benchmark providing parallel corpora across diverse PLs with accompanying test suites for validation.
Early examples include \textit{CodeNet}~\cite{Puri2021CodeNet}, \textit{AVATAR}~\cite{Ahmad2022AVATAR}, \textit{G-TransEval}~\cite{Jiao2024GTransEval}, and \textit{TransCoder-test} ~\cite{Roziere2020TransCoder,yang2024exploring}, sourced from programming contest sites. 
To further extend the evaluation scope and complexity, \textit{CodeTransOcean}~\cite{Yan2024CodeTransOcean} offers a large-scale multilingual dataset of popular and niche PLs with over 7,000 samples, although only 44 algorithmic tasks simultaneously cover all parallel PLs~\cite{xue2024classeval-t}.
\textit{xCodeEval}~\cite{Lu2023xCodeEval} is an execution-based, multilingual benchmark derived from Codeforces \cite{Codeforc25:online} supporting both code generation and translation tasks. \textit{PolyHumanEval}~\cite{Tao2023PolyHumanEval} extends \textit{HumanEval}~\cite{Chen2021HumanEval} to 14 PLs, and \textit{ClassEval-T}~\cite{xue2024classeval-t} introduced class-level translation tasks, incorporating field, method, and library dependencies.

Later on, benchmarks addressing real-world repository-level translation have been proposed. Ibrahimzada et al. \cite{ibrahimzada2024repository} and \textit{RepoTransBench}~\cite{wang2024repotransbench} evaluate LLMs on entire repositories but proactively exclude TPLs to alleviate translation difficulties. \textit{RustRepoTrans}~\cite{ou2024repository} is the first function-level code translation benchmark targeting Rust with the context of corresponding repositories. Although comprising 375 tasks, it lacks parallel corpora across all PLs, limiting its evaluation utility for cross-lingual generalization and transfer.
\textit{C2R-Bench}~\cite{wang2025evoc2rust}, Gao et al.~\cite{gao2024rule}, and Zhou et al.~\cite{zhou2025porting} concentrate on domain-specific PLs migration, including Rust for system programming and Swift/ArkTS for mobile applications, significantly limiting the coverage of TPL categories.
Table~\ref{benchmark} provides a comparative overview of these benchmarks against \textit{TransLibEval}, detailing their supported PLs, practical alignment, granularity, task amount (\textbf{\#Task}), test amount (\textbf{\#Tests/T}), sample scales (\textbf{\#LOC/T} and \textbf{\#Tokens/T}), and TPL amount (\textbf{\#LD})/percentage (\textbf{LD\%}). Considering the inherent flaws of non-parallel benchmarks mentioned in Xue et al. \cite{xue2024automated}, which always happen at the repository level, only benchmarks with parallel PLs are included for comparison. As can be seen, (1) most benchmarks carry limited code scale and complexity (e.g., LOC of 14.3 for \textit{G-TransEval}, 9.6 for \textit{PolyHumanEval} on average). Except \textit{ClassEval-T}, a class-level benchmark, \textit{TransLibEval} surpasses others by 88\% on average. (2) Although \textit{TransLibEval} contains only 5 test cases per task, lower than most counterparts, its statement coverage can reach 95\%, significantly saving the evaluation cost. (3) More importantly, the proportion of tasks involving library dependencies (\textbf{LD\%}) is strikingly low across most benchmarks (e.g., 10.5\% in \textit{CodeNet}, 6.8\% in \textit{CodeTransOcean}, and 4.3\% in \textit{PolyHumanEval}). \textit{ClassEval-T} contains a relatively high percentage of library dependencies, but the scale is limited to 74, which is less than half of \textit{TransLibEval}. In contrast, \textit{TransLibEval} includes 200 libraries across 10 categories (shown in Table \ref{tab:third_party_libraries}), making it an effective code translation benchmark to fill the significant gap of previous related studies. 

\begin{table}[t]
\centering
\caption{Comparative Analysis of Code Translation Benchmarks}
\vspace{-0.4cm}
\label{benchmark}
\scriptsize
\setlength{\tabcolsep}{2.0pt}
\renewcommand{\arraystretch}{1.1}
\resizebox{\textwidth}{!}{
\begin{tabular}{@{}lccccccccccc@{}}
\toprule
\textbf{Benchmark} & \textbf{Year} & \textbf{Parallel Programming Languages} & 
\textbf{Practical}  & \textbf{Granularity} &\textbf{\#Tasks} & \textbf{\#Tests/T} & \textbf{\#LOC/T} & \textbf{\#Tokens/T} & \textbf{\#LD} & \textbf{LD\%} \\ 
\midrule
\textit{CodeNet} \cite{Puri2021CodeNet} & 2021 & C,C++,Go,Java,Python  & N          & Statement/Method-level & 200 & 1.0 & 34.9 & 112.0 & 21 & 10.5\% \\
\textit{AVATAR} \cite{ahmad2023avatar} & 2021 & Java,Python   & N          & Statement/Method-level & 250 & 25.1 & 26.9 & 149.5 & 51 & 20.4\% \\
\textit{G-TransEval} \cite{jiao2023evaluation} & 2023 & C++,C\#,Java,Python,JavaScript                 & N          & Method-level      & 400 & 5.0 & 14.3 & 93.6 & - & - \\
\begin{tabular}[l]{@{}l@{}}\textit{CodeTransOcean-}\\ 
\textit{MultilingualTrans} \cite{yan2023codetransocean}\end{tabular} & 2023 &  \begin{tabular}[c]{@{}c@{}}C,C++,C\#,Java,Python,\\ Go,PHP,Visual Basic\end{tabular}   & N          & Statement/Method-level & 44 & 1.0 & 23.8 & 82.1 & 3 & 6.8\% \\
\textit{xCodeEval} \cite{khan2024xcodeeval} & 2024 & \begin{tabular}[c]{@{}c@{}}C,C++,C\#,Java,Python,Ruby,Go,\\ JavaScript,Kotlin,PHP,Rust\end{tabular} & N          & Statement/Method-level & 226 & 46.8 & 30.8 & 95.5 & 27 & 11.9\% \\
\textit{Yang} et al. \cite{yang2024exploring} & 2024 & Python,Java,C++  & N          & Method-level& 568 & 6.2 & 12.4 & 95.8 & - & - \\
\textit{PolyHumanEval} \cite{tao2024unraveling} & 2024 & \begin{tabular}[c]{@{}c@{}}Python,C++,C\#,Dart,Go,Java,JavaScript,PHP,\\ Kotlin,Ruby,Rust,Scala,Swift,TypeScript\end{tabular} & N          & Method-level& 164 & 8.1 & 9.6 & 37.9 & 7 & 4.3\% \\
\textit{ClassEval-T} \cite{xue2024classeval-t} & 2025 & Python,Java,C++  & Y & Class-level& 94 & 33.8 & 66.7 & 199.5 & 74 & 78.7\% \\
\addlinespace[0.5em]
\rowcolor{gray!10}
\textbf{\textit{TransLibEval} (Ours)} & \textbf{2025} & \textbf{Python,Java,C++} & \textbf{Y} & \textbf{Method-level}& \textbf{200} & \textbf{5.0} & \textbf{41.0} & \textbf{120.0} & \textbf{200} & \textbf{100\%} \\
\bottomrule
\end{tabular}
} 
\vspace{-1mm}
\begin{minipage}{\textwidth}
\footnotesize
\vspace{2pt}
\raggedright
\textbf{Note:} \#Tests/T = Test case number per task; \#LOC/T = Line number of code per task; 
\#Tokens/T = Code token number per task; \#LD = Task number with library dependencies; 
LD\% = Percentage of tasks with dependencies.
\end{minipage}
\end{table}

\vspace{-0.9em}
\section{New Benchmark: \textit{TransLibEval}}




In the code translation field, a multi-PL parallel benchmark is always constructed from a calibration PL, based on which to extend to other counterpart PLs manually \cite{yang2024exploring, xue2024classeval-t,Tao2023PolyHumanEval}. Considering the extensive Python third-party libraries and their huge community support, we select Python as a calibration PL for benchmark construction, thereby facilitating the library selection and debugging. Additionally, following most code translation settings \cite{yang2024exploring, xue2024classeval-t, Jiao2024GTransEval,ibrahimzada2024repository}, C++ and Java are chosen as the counterpart PLs in this work. We detail the construction of the calibration and counterpart portion of \textit{TransLibEval}, respectively.

\subsection{Calibration Construction}
To construct a Python calibration, we first identify Python libraries based on their usage frequency in PyPI download statistics \cite{flynn2024pypi} (2020-2024), such as data processing (e.g., numpy, pandas), machine learning (e.g., scikit-learn, lightgbm), web development (e.g., requests, beautifulsoup4), and visualization (e.g., matplotlib, seaborn). For each of the above Python libraries, we formulate four method-level tasks to implement its most typical functionalities. In addition, we select the method-level granularity because it can isolate most external influences (e.g., cross-file/class/method dependencies), making our experiments exclusively focused on LLMs' capability on TPL-targeted code translation. To make the whole implementation more canonical, we follow a series of principles according to previous studies \cite{xue2024classeval-t,du2024evaluating}. \textbf{(1) For naming conventions.} We follow an officially recommended Python style guide (\textit{PEP 8} \cite{PEP8–Sty81:online}) to name identifiers, where methods and variables are named in snake\_case (e.g., \texttt{compute\_area}), classes are required to use the pascal case (e.g., \texttt{AreaCalculator}), and constants follow the screaming snake case style (e.g., \texttt{MAX\_ITERATIONS}). \textbf{(2) As for the implementation layout.} Considering that the follow-up translation involves Java programs, whose methods cannot be independently executed without a wrapped class. We require that each Python task must define exactly one top-level class encapsulating a single instance method calling a designated third-party API to exercise library-specific operations, thereby making the parallel corpora more consistent across coding styles. Besides, we use 4 spaces per indentation level according to \textit{PEP 8}, and comments are excluded to avoid additional hints for LLMs when translation. \textbf{(3) Towards type conventions.} Following Du et al. \cite{du2024evaluating}, we require that all method parameters and return values shall be restricted to primitive data types (\texttt{int}, \texttt{float}, \texttt{boolean}, \texttt{string}, etc.). Composite/Object-level data types (e.g., matrices, custom objects) are explicitly prohibited. This constraint guarantees that when invoking focal methods externally from test suites, only limited data types need to be considered across diverse PLs, facilitating the test suite construction and the whole evaluation procedure. \textbf{(4) For test suite construction.} We employ Python's \textit{unittest} framework \cite{unittest} to construct method-level test cases, utilizing its comprehensive assertion APIs and test fixtures such as \texttt{setUp()} for pre-test initialization to preserve five essential test categories, including normal operation with valid inputs, edge case evaluation, exception handling scenarios, input type validation, and resource constraint testing. 
Adhering to the above established principles, two software engineers with $\geq$ 3 years of professional expertise in Python, Java, and C++ accomplish the above calibration construction, where one is responsible for the specific implementation, while the other's duty is code review and correctness verification. The detailed construction procedures along with a full-workflow example are provided in our repository for reference.\cite{TransLib77:online}
Finally, we select 53 libraries to implement their typical functionalities, obtaining 212 code samples, which cost a total of 300 person-hours. The specific statistics are shown in Table \ref{tab:third_party_libraries}.

\begin{table}[t]
\centering
\caption{The Categorization of Studied Third-Party Libraries for Python, Java, and C++}
\vspace{-0.4cm}
\label{tab:third_party_libraries}
\tiny
\setlength{\tabcolsep}{1.5pt}
\renewcommand{\arraystretch}{1.2}

\begin{tabularx}{\textwidth}{@{}p{1.6cm} p{5.5cm} p{2.2cm} p{2cm} p{1.5cm} c@{}}
\toprule
\textbf{Category} & \textbf{Description} & \textbf{Examples-Python} & \textbf{Examples-Java} & \textbf{Examples-C++} & \textbf{\#Tasks} \\
\midrule
Data Processing & Numerical computation, and algorithm implementation. & numpy, pandas & JDBI, Jackson & boost, Eigen & 68 \\
Utilities & Tools for common utilities, formatting, and extensions. & tqdm, Guava & Commons IO & boost & 36 \\
Machine Learning & Libraries supporting ML models' training and inference. & scikit-learn,{\quad}lightgbm & deeplearning4j, Smile & xgboost, fasttext & 28 \\
Web Development & Tools for networking and web application development. & requests, beautifulsoup4 & OkHttp, Spring Context & libcurl & 24 \\
Visualization & Libraries for plotting charts and visualizing data. & matplotlib, seaborn & JFreeChart & plplot & 20 \\

NLP & Libraries for text processing, linguistic analysis, and NLP tasks. & nltk & Stanford CoreNLP & cryptopp & 20 \\

Graphics & Libraries for image processing and computer vision. & opencv, PIL & Lucene & opencv & 8 \\
Database & Tools for database connection, ORM, and SQL operations. & peewee & JDBI & sqlite3 & 4 \\
Security & Encryption, hashing, and security-related functionalities. & pycrypto & BouncyCastle & cryptopp & 4 \\
Other & Miscellaneous libraries that do not fit into the above categories. & jsonschema & Commons Lang & - & 4 \\
\bottomrule
\end{tabularx}
\vspace{-0.55cm}
\end{table}

\subsection{Counterpart Construction}
To construct the counterpart portion of \textit{TranLibEval} based on the above calibration, i.e., manually translate the above Python corpus to Java and C++ sides, we strictly follow the five principles established by Xue et al. \cite{xue2024classeval-t} to carry out the whole translation process. For example, the naming convention (Principle 1) and implementation layout (Principle 3) of translated Java and C++ programs follow their officially recommended style guides \cite{reddy2000java, sutter2004c++}. In the type convention aspect (Principle 2), they also defined specific data type mapping rules between Python and Java/C++. From the library selection (Principle 4) and test suite construction (Principle 5) perspectives, explicit library searching workarounds, library management tools (\textit{maven} \cite{maven} for Java and \textit{nuget} \cite{nuget} for C++), and testing frameworks (\textit{JUnit} \cite{junit5} for Java and \textit{GoogleTest} \cite{googletest} for C++) are also designated.

Specifically, five software engineers with $\geq$ 3 years of professional expertise in Python, Java, and C++ are involved. The lead architect assumed ultimate responsibility for review and arbitration, while the remaining members formed two PL-specific expert groups. Each group is responsible for either Java or C++-oriented translation and is composed of two members, one is 
a primary translator and the other is a dedicated verifier. The primary translator adheres to the five principles established by Xue et al. to translate Python to Java/C++, including both the TPL-related programs and their corresponding test suites. Afterwards, for each translation sample, the verifier first exhaustively reviews the translated result, examining whether it conforms to the five principles, and then executes these test cases to double-check the correctness of the translation. Among the five principles, disagreements between translators and verifiers normally arise in type handling (Principle 2), library selection (Principle 4), and test equivalence (Principle 5). At this point, the final decision is made by lead architect. After the aforementioned manual translation process, 12 samples still cannot reach a consensus, even with the engagement of the lead architect, owing to their translation difficulties. For example, \textit{spaCy} and \textit{TensorFlow} are two of this kind, where the former lacks any Java/C++ equivalent while the latter exhibits significant API divergence across PL bindings. We exclude these 12 samples and finally yield a parallel TPL-targeted code translation benchmark with 200 samples. The entire construction consumed 200 person-hours.

\section{Experimental Design}
This study utilizes the \textit{TransLibEval} benchmark to evaluate recent LLMs on library-centric code translation. We aim to address the following research questions (RQs):

\textbf{RQ1 (Overall Correctness): How do recent LLMs perform on library-centric code translation?} Compared with general code translation, library-centric translation introduces additional challenges due to the need for correct recognition, import, and use of third-party libraries. Given the increasing prevalence of external library APIs in modern software development, it becomes critical to understand how well LLMs can handle such dependencies across languages.

\textbf{RQ2 (Translation Strategies): How do different translation strategies affect the performance of recent LLMs on library-centric code translation?} Developers may adopt different prompting strategies like direct translation, Intermediate Representation (IR)-guided translation, or retrieval-augmented translation. Evaluating how these strategies impact the handling of library dependencies can offer practical guidance for prompt engineering and model deployment.

\textbf{RQ3 (Library Dependency Awareness): How do the LLMs perform in identifying necessary and available libraries?}
Library-centric code translation often requires not just syntax conversion but also awareness of the necessary libraries for implementing equivalent functionalities. Considering that the library mappings between different PLs are not always one-to-one, LLMs have to infer a limited library set to accomplish the translation task in this work. Nonetheless, LLMs' above-mentioned ability is still unknown, and the corresponding investigation has not been extensively conducted before, thus motivating us to delve into this RQ.


\textbf{RQ4 (Failed Cases Analysis): What kind of errors do LLMs make in library-centric translation, and how frequent are they?} While prior work has examined LLM failures in method-, class- or even repo-level translation \cite{yang2024exploring,tao2024unraveling,pan2024lost,ou2024repository,xin2025enhancing}, their included libraries, even in repo-level, are limited as shown in Section \ref{Related Work}. Thus, a large-scale fine-grained analysis of library-related errors is still lacking, and identifying these failure modes can help the community develop more targeted improvements for LLM-based code translation systems.

\subsection{Studied LLMs}
{
Table \ref{tab:studied_llms} lists our studied LLMs with their categories, base model, released times (Time), parameter sizes (Size), whether they utilize Mixture-of-Experts (MoE) and Rotary Position Embedding (RoPE) techniques, pretraining token numbers (Training Base), and the input/output context window token limits (In/Out). For those models whose context windows are measured in characters, we specified them separately.} {As can be seen, to extensively explore the performance of the latest LLMs, we select LLMs from general (i.e., Llama 3 \cite{Llama3268:online} and Gemma \cite{team2024gemma}), code (i.e., CodeLlama \cite{roziere2023code}), and commercial kinds 
\footnote{Following \cite{xue2024classeval-t}, commercial LLMs are defined as large language models with substantial parameter sizes  (typically exceeding 100B), developed by organizations for commercial purposes, and providing publicly accessible interfaces for interactive dialogue and text-based applications.}
(i.e., DeepSeek-V3 \cite{DeepSeek19:online}, GPT-4o \cite{OpenAI59:online}, GPT-3.5-Turbo and Qwen-Max \cite{qwen2025}), released from late 2023 to late 2024.} Furthermore, for LLMs from the same family, we also select two different sizes (i.e., Llama3-8B/70B) to study the influence of the volume of model parameters.

\begin{table}[htbp]
\centering 
\setlength\tabcolsep{6.4pt}
\caption{Studied LLMs in this paper}
\vspace{-0.9em}
\label{tab:studied_llms} 
\tiny
\begin{tabular}{lcccccccc} 

\toprule
Category & Model & Base Model & Time & Size & MoE & RoPE & Training Base (Tokens) & In/Out (Tokens)  \\
\midrule
\multirow{4}{*}{Commercial}
    & DeepSeek-V3 \cite{DeepSeek19:online} & - & 2024-12 & 671B & Yes & Yes & 14.8 trillion & 64k / 4096  \\
    & GPT-4o \cite{OpenAI59:online}& - & 2024-05 & $\sim$200B \cite{abacha2024medec} & Yes & Yes & - & 128k / 4096 \\
    & GPT-3.5-Turbo & - & 2023-03 & - & No & Yes & - & 4k / 4096 \\
    & Qwen-Max \cite{qwen2025} & - & 2025-01 & - & Yes & Yes & 20 trillion & 30k / 8192 \\
\midrule
\multirow{2}{*}{General}
    & Llama3 \cite{Llama3268:online} & - & 2024-04 & 8B & No & Yes & 15.6 trillion & 7k / 1024\\
    & Llama3 \cite{Llama3268:online} & - & 2024-04 & 70B & No & Yes & 15.6 trillion & 7k / 1024 \\
\midrule
Code
    & CodeLlama \cite{roziere2023code} & Llama2 \cite{touvron2023llama} & 2023-08 & 7B & No & Yes & 500 billion & 8000 Characters/ 1024 \\
\bottomrule
\end{tabular}

\begin{tablenotes} 
\definecolor{light cyan}{HTML}{E1FFFF} 
\fboxsep1.5pt
\item \scriptsize {In the following sections, ``DeepSeek'' refers to DeepSeek-V3, ``Qwen'' indicates Qwen-Max, and ``GPT-3.5'' indicates GPT-3.5-Turbo. We also use ``Llama3-70B'' and ``Llama3-8B'' to denote the two versions of Llama3.}

\end{tablenotes}
\vspace{-2em} 
\end{table}


\subsection{Studied Translation Strategies}
\label{Studied Translation Strategies}

Given a library-centric code translation task, we investigate the following six translation strategies of three categories for each of the studied LLMs.

\textbf{(1) Direct Translation (Direct):} LLMs are provided with the raw source-language code snippet involving third-party library usage and are asked to translate it into the target PL directly. This strategy simulates the most straightforward application of LLMs in real-world scenarios where developers expect end-to-end translation without any auxiliary reformulation or decomposition of the input. A prompt example is shown in \cite{TransLib77:online}.

\textbf{(2) IR-Guided Translation (IR):} 
IRs, such as pseudocode, reasoning traces, or summaries, have been widely used in LLM-based code intelligence \cite{SCoT,jiang2024self,yuan2024evaluating}. Its core idea is requiring LLMs to generate an IR first, as a more explicit and specific instruction, to guide them in producing the final code snippets. Apart from the above, we consider that IR also has the potential to isolate the influence of source PLs, making LLMs focus more on the functionality implementation. This is a critical motivation in library-centric code translation, as many previous studies \cite{yang2021multi,pan2024lost} have revealed that LLMs tend to copy APIs of source PLs, which could be more severe in our library-intensive setting. Below, we introduce three typical IR-based prompting methods for this study. Prompt examples of each variant are shown in \cite{TransLib77:online}.

\begin{itemize}
\item[$\bullet$] \emph{IR(CoT)}: It first requires LLMs to break down the program execution logic of source PL into a sequence of natural language reasoning steps. Then, LLMs are guided to translate the program into the target PL version based on the above IR prompt.

\item[$\bullet$] \emph{IR(pseudocode)}: It replaces the above natural language reasoning sequences with pseudocode as the IR between source and target PLs, where the pseudocode consists of essential program elements, such as branches, loops, and hierarchical structures.

\item[$\bullet$] \emph{IR(summary)}: It first requires LLMs to generate a summary for the program to be translated, where the summary contains the core functionality and input/output requirements. Subsequently, LLMs are prompted to translate the program of the source PL to its counterpart in the target PL based on the above summary.


\end{itemize}

\textbf{(3) Retrieval-Augmented Translation (RA):} 
Owing to the complex library mapping relation among diverse PLs, previous studies \cite{pan2024lost,xue2024classeval-t} have noticed that correctly identifying and applying necessary libraries in the target PL when translating code is critically difficult for current LLMs. To address this challenge, an intuitive direction is to inject external knowledge into LLMs via retrieval, thus RA becomes another kind of highly motivated method to be studied in library-centric code translation. Specifically, we retrieve relevant external knowledge from Stack Overflow \cite{stackoverflow} to augment LLMs' capability. We explore two variants, and their prompt examples are shown in \cite{TransLib77:online}.
\begin{itemize}
\item[$\bullet$]\emph{RA(method)}: It instructs LLMs to analyze the entire function and generate a dedicated search query for code translation.  Subsequently, based on Stack Exchange API \cite{StackExc69:online}, we develop a retrieval tool that searches on Stack Overflow with the above-generated query, and returns the top-\textit{k} most relevant posts. For each post, the retrieval tool fetches the top-\textit{n} most upvoted answers for prompting LLMs to identify specific library/API usage and generate the final translation results. We set \textit{k}=1 and \textit{n}=3 in this study.


\item[$\bullet$]\emph{RA(name)}: Method names normally convey program functionalities more succinctly compared with RA(method)'s query. As an alternative to \emph{RA(method)}, \emph{RA(Name)} uses only the target PLs plus method names (e.g., ~\texttt{Python: "image\_to\_grayscale"}) 
as queries to retrieve external knowledge. Other settings and procedures are all the same as \emph{RA(method)}.


\end{itemize}


\subsection{Evaluation Metrics}

\label{Evaluation Metrics}

We follow previous studies \cite{roziere2021leveraging,roziere2020unsupervised,yang2024exploring} and evaluate LLMs via Compilation Success Rate (CSR), Pass Rate (PR), Computational Accuracy (CA), and Library Dependency Awareness rate (LDA).

\textbf{Compilation Success Rate (CSR):} It computes the ratio of samples that can be successfully compiled after translation. We define CSR as below.

\vspace{-0.2cm}

\begin{equation}
    CSR = \frac{\sum^{N}_{k=1}cs(\hat{y_{k}})}{N}, \ \textbf{where} \ cs(\hat{y_{k}}) = \left\{\begin{matrix}
1 & Compile(\hat{y_{k}}) \to success\\
0 & Compile(\hat{y_{k}})  \to error\\
\end{matrix}\right. 
\end{equation}

where $N$ denotes the total number of samples, $\hat{y_{k}}$ denotes the $k$-th translated sample via a certain LLM. \textit{Compile($\cdot$)} denotes compiling samples with their corresponding compilers, such as \textit{javac} for Java while \textit{g++} for C++, and its results are either \textit{success} or \textit{error}.

\textbf{Pass Rate (PR):} It computes the ratio of successfully passed test cases across all samples, where a passed test case means the execution result of a translated program equals that of the ground truth program. We define PR as below.

\vspace{-0.2cm}
\begin{equation}
PR = \frac{\sum^{N}_{k=1}\sum^{T_{k}}_{j=1}ps(\hat{y_{k}},y_{k}, j)}{\sum^{N}_{k=1}T_{k}}, \ \textbf{where} \ ps(\hat{y_{k}},y_{k}, j) = \left\{\begin{matrix}
1 & Exec_{k,j}(y_k,j)=Exec_{k,j}(\hat{y_{k}},j)\\
0 & Exec_{k,j}(y_k,j)\neq Exec_{k,j}(\hat{y_{k}},j) \\
\end{matrix}\right. 
\end{equation}

where $T_{k}$ is the total number of test cases for the $k$-th sample, and $y_k$ denotes the ground truth of the $k$-th sample. $Exec_{k,j}(\cdot)$ denotes the execution result of a program ($y_k$ or $\hat{y_k}$) on the $j$-th test case from the test suite of the $k$-th sample.

\textbf{Computational Accuracy (CA):} It computes the ratio that the translated programs can produce the same execution result as the ground truths, given a set of test suites. We define CA as below.

\vspace{-0.2cm}
\begin{equation}
CA = \frac{\sum^{N}_{k=1}ca(y_{k},\hat{y_{k}})}{N}, \ \textbf{where} \ ca(y_{k},\hat{y_{k}}) = \left\{\begin{matrix}
1 & Exec_{k}(y_k)=Exec_{k}(\hat{y_{k}})\\
0 & Exec_{k}(y_k)\neq Exec_{k}(\hat{y_{k}}) \\
\end{matrix}\right. \label{eq2}
\end{equation}

where 
$Exec_{k}(\cdot)$ denotes the execution result of a program with the test suite of the $k$-th sample.  

\textbf{Library Dependency Awareness Rate (LDA):} To assess whether LLMs can correctly identify useful third-party libraries when translation, we adapt the dependency recall metric from prior works \cite{du2024evaluating,yu2024codereval}. Specifically, LDA measures the ratio of translated programs that invoke useful third-party libraries to fully achieve the intended functionalities, which can be formulated as below. 


\vspace{-0.2cm}
\begin{equation}
LDA = \frac{N_{lib}}{N}
\end{equation}

where $N_{lib}$ denotes the number of translated samples that invoke useful libraries. We delineate the specific definition of invoking useful libraries and the counting methodology of $N_{lib}$ in Section \ref{RQ3: Library-Dependency Awareness}.

\subsection{Implementation Details}

We utilize seven LLMs for our experiments. Specifically, we employ the Deepseek, Qwen-Max, GPT-4o, and GPT-3.5 via their official API interface\cite{deepseek,OpenAI59:online}. Additionally, CodeLlama, Llama3-8B, and Llama3-70B are operated via the open API interface provided by Baidu Qianfan platform\cite{qianfan}. The hyperparameters for the code translation process are set as follows: $temperature = 0$ and $n = 1$, which means we only fetch LLMs' first translated candidates for evaluation during inference, and the LLMs' randomness can be eliminated to a minimum, thereby ensuring the reproducibility of the experiments. Nonetheless, despite this, the randomness persists to some extent. 
Thus, we assessed stability by repeating a subset of experiments three times following previous studies \cite{tip2025llmorpheus}.  This subset specifically comprised C++ to Python translations using the Direct strategy. The results showed that all LLMs except GPT-3.5-Turbo exhibited a Pass Rate variance of less than 5\%, indicating consistently stable behavior. GPT-3.5-Turbo showed a deviation of 5.9\%. The overall low variance observed in this analysis suggests that the inherent randomness had a negligible influence on the reliability of our conclusions for the study as a whole. All other hyperparameters are kept by default. All evaluations are performed in a one-shot setting.


\section{Experimental Results}
\label{Experimental Results}
\subsection{RQ1: Overall Correctness}
\label{RQ1: Overall Correctness}
To intuitively understand the difficulty and motivation of library-centric code translation tasks, we evaluate each studied LLM on both the \textit{TransLibEval} and the benchmark released by Yang et al.\cite{yang2024exploring}, where the latter is also a method-level code translation benchmark but TPL-free, thereby making a direct comparison. 
Figure \ref{rq1fig} demonstrates their correctness comparison in terms of average Compilation Success Rate (CSR), Computational Accuracy (CA), and Pass Rate (PR). 
Besides, Table \ref{rq1table} presents the detailed experimental results on \textit{TransLibEval}. It should be noted that the Direct translation strategy is used for LLMs in RQ1, and we have the following observations.

\textbf{Library-free vs. Third-party library-centric Translation.} As shown in Figure \ref{rq1fig}, the code translation performance of LLMs degenerates significantly and consistently on all translation pairs when converting from library-free to TPL-centric benchmark. Specifically, all LLMs studied can achieve CSR over 95\%, CA over 78\% and PR over 76\% on almost all translation pairs in Yang et al.'s benchmark. However, when assessing them on the \textit{TransLibEval}, LLMs' code translation performance declines by 41.63\% on average in terms of CSR, 61.03\% on average in terms of CA, while by 54.49\% on average in terms of PR. The most challenging translation pairs, Python-to-Java and Python-to-C++, suffer catastrophic declines of 61.92\% and 63.16\% in CSR, 82.96\% and 73.48\% in CA, and 75.83\% and 76.24\% in PR, respectively. The dramatic performance degradation can be attributed to the inherent complexity of involving practical TPLs in code translation, which involves handling intricate API dependencies, complex type systems, and sophisticated library-specific semantics that are absent in library-free code translations. 

\begin{figure*}[htbp]
  \vspace{-1.5em}
  \setlength{\abovecaptionskip}{0pt}
  \centering
  \includegraphics[width=0.8\textwidth]{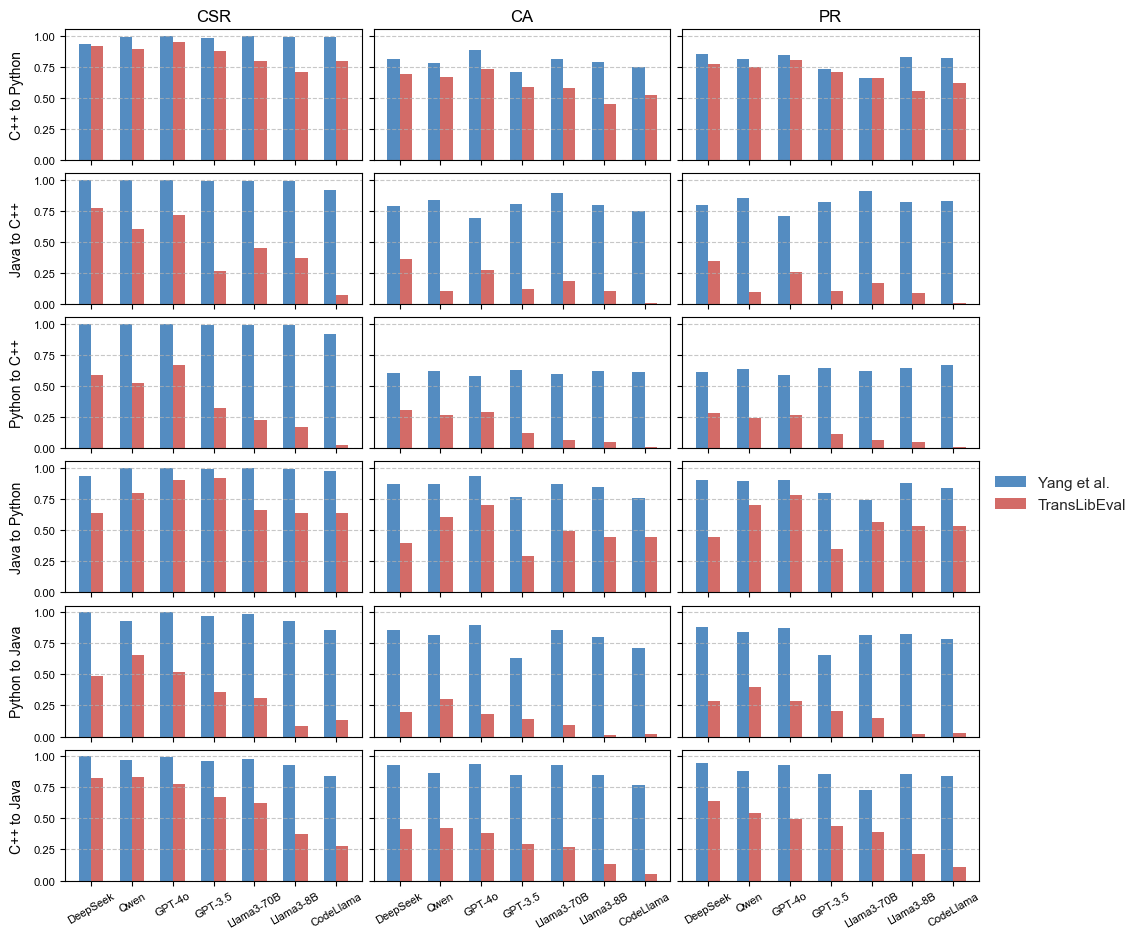}
  \caption{Performance Comparison between \textit{TransLibEval} and Yang et al. among Diverse LLMs}
    \label{rq1fig}
\end{figure*}

\begin{boxK}
\small \faIcon{pencil-alt} \textbf{Finding 1:} Compared with method-level code translations, LLMs show a marked decline in performance when translating code involving TPLs, with Python-to-Java and Python-to-C++ translations suffering the most severe degradation (e.g., over 80\% in CA), primarily due to the complex API dependencies and library-specific semantics.
\end{boxK}

\textbf{Comparison among LLMs.} Focusing on the results of the \textit{TransLibEval} benchmark (see Table \ref{rq1table}), commercial LLMs (GPT-4o, GPT-3.5, DeepSeek, and Qwen) consistently perform the best, achieving average scores of 68.56\% in terms of CSR, 36.94\% in terms of CA, and 43.05\% in terms of PR across various translation pairs. Commercial LLMs demonstrate a significant advantage over smaller LLMs, with relative improvements of 67.79\% in CSR, 68.11\% in CA, and 62.59\% in PR. Among commercial LLMs, GPT-4o achieves the highest performance with 75.42\% in terms of CSR, 42.67\% in terms of CA, and 48.28\% in terms of PR, followed closely by DeepSeek and Qwen. This superiority can be attributed to their more sophisticated architecture, larger parameter sizes, and extensive training on diverse code corpora that include TPL usage patterns. Among smaller LLMs, Llama3-70B demonstrates the strongest overall performance, yet it still lags considerably behind leading commercial counterparts. Notably, compared with Llama3-8B, CodeLlama performs weakly overall, although their parameter sizes are almost the same, and the latter was pre-trained on code. We infer the reason is: CodeLlama is built upon the Llama2 \cite{touvron2023llama}, whose training base is 2 trillion tokens, which only accounts for one seventh of Llama3-8B's training base, as shown in Table \ref{tab:studied_llms}.


\begin{boxK}
\small \faIcon{pencil-alt} \textbf{Finding 2:} 
{Commercial LLMs exhibit predominantly superior performance on code translation involving TPLs, significantly outperforming smaller LLMs by 67.79\% in CSR, 68.11\% in CA and 62.59\% in PR. LLMs with limited training bases or smaller sizes always perform more weakly. 

}
\end{boxK}

\textbf{Comparison among translation directions.} Based on Figure \ref{rq1fig} and Table \ref{rq1table}, LLMs generally perform much better on Python-oriented translations (e.g., C++/Java-to-Python) in terms of all evaluation metrics. Python-oriented translations achieve average scores of 54.39\% in terms of CA, 79.46\% in terms of CSR, and 62.64\% in terms of PR, representing substantial advantages of 192.60\%, 75.40\%, and 177.11\% in terms of each metric in order compared to other translation directions. However, concerning translations to C++ and Java, LLMs' performance drops dramatically, particularly, Python-to-C++ and Python-to-Java show the most significant performance degradation, with CSR falling below 37\%, CA falling below 17\% and PR below 15\%. 
The performance differences can be attributed to both pre-training data biases and the inherent complexity of different PLs. Firstly, the superior performance in Python-oriented translations aligns with Python's dominance in open-source version control platforms (e.g., GitHub) since 2022 \cite{Thetoppr42:online}, which are the most prevalent code databases for LLMs' pre-training \cite{zhao2024deciphering, liu2024deepseek, team2024codegemma, roziere2023code, tao2024unraveling}. Secondly, Python, as an interpreted PL with a dynamic type system, carries more human-readable syntax in library usage, making it analogous to natural language and facilitating LLMs' learning \cite{wentworth2011think}. 
In contrast, LLMs face greater challenges when dealing with the strict type systems, complex memory management requirements, and intricate library interfaces of C++ and Java in third-party library contexts \cite{prechelt2000empirical}.

\begin{boxK}
\small \faIcon{pencil-alt} \textbf{Finding 3:} 
LLMs manifest substantial superiority in translating other PLs to Python within TPL contexts, while struggling significantly with translations to C++ and Java (e.g., 54.39\% v.s. 18.59\% in CA) due to both the training base and language characteristics.
\end{boxK}

\begin{table}[]
\vspace{-1em}
\setlength{\abovecaptionskip}{0cm}
\caption{Experimental Results with direct translation on \textit{TransLibEval}}
\label{rq1table}
\tiny
\setlength\tabcolsep{2pt}
\begin{threeparttable}
\begin{tabular}{lccccccccccccccccccccc}
\toprule
\multirow{2.4}{*}{\textbf{Models}} & \multicolumn{3}{c}{\textbf{C++ to Python}} & \multicolumn{3}{c}{\textbf{Python to C++}} & \multicolumn{3}{c}{\textbf{Java to Python}} & \multicolumn{3}{c}{\textbf{Python to Java}} & \multicolumn{3}{c}{\textbf{C++ to Java}} & \multicolumn{3}{c}{\textbf{Java to C++}} & \multicolumn{3}{c}{\textbf{Average}} \\
\cmidrule(lr){2-4} \cmidrule(lr){5-7} \cmidrule(lr){8-10} \cmidrule(lr){11-13} \cmidrule(lr){14-16} \cmidrule(lr){17-19} \cmidrule(lr){20-22}
& \multicolumn{1}{c}{CSR} & \multicolumn{1}{c}{CA} & \multicolumn{1}{c}{PR} & \multicolumn{1}{c}{CSR} & \multicolumn{1}{c}{CA} & \multicolumn{1}{c}{PR} & \multicolumn{1}{c}{CSR} & \multicolumn{1}{c}{CA} & \multicolumn{1}{c}{PR} & \multicolumn{1}{c}{CSR} & \multicolumn{1}{c}{CA} & \multicolumn{1}{c}{PR} & \multicolumn{1}{c}{CSR} & \multicolumn{1}{c}{CA} & \multicolumn{1}{c}{PR} & \multicolumn{1}{c}{CSR} & \multicolumn{1}{c}{CA} & \multicolumn{1}{c}{PR} & \multicolumn{1}{c}{CSR} & \multicolumn{1}{c}{CA} & \multicolumn{1}{c}{PR} \\\midrule
DeepSeek & 91.50 & 69.50 & 77.60 & 59.00 & \cellcolor[HTML]{E1FFFF}31.00 & \cellcolor[HTML]{E1FFFF}28.60 & 64.00 & 40.00 & 44.50 & 48.50 & 20.00 & 28.70 & 82.00 & 41.00 & \cellcolor[HTML]{E1FFFF}64.00 & \cellcolor[HTML]{E1FFFF}77.00 & \cellcolor[HTML]{E1FFFF}36.50 & \cellcolor[HTML]{E1FFFF}34.70  & 70.33 & \underline{39.67} & \underline{46.35} \\
Qwen & 89.00 & 67.00 & 75.20 & 52.50 & 27.00 & 24.70 & 80.00 & 60.50 & 69.90 & \cellcolor[HTML]{E1FFFF}65.50 & \cellcolor[HTML]{E1FFFF}30.00 & \cellcolor[HTML]{E1FFFF}39.50 & \cellcolor[HTML]{E1FFFF}83.00 & \cellcolor[HTML]{E1FFFF}42.00 & 54.30 & 60.50 & 10.50 & 9.60  & \underline{71.75} & 39.50 & 45.53 \\
GPT-4o & \cellcolor[HTML]{E1FFFF}94.50 & \cellcolor[HTML]{E1FFFF}73.00 & \cellcolor[HTML]{E1FFFF}80.80 & \cellcolor[HTML]{E1FFFF}66.50 & 29.00 & 27.10 & 90.50 & \cellcolor[HTML]{E1FFFF}70.00 & \cellcolor[HTML]{E1FFFF}78.20 & 52.00 & 18.00 & 28.01 & 77.50 & 38.50 & 49.50 & 71.50 & 27.50 & 26.00  & \textbf{75.42} & \textbf{42.67} & \textbf{48.27} \\
GPT-3.5 & 87.50 & 58.50 & 70.40 & 32.50 & 12.50 & 11.90 & \cellcolor[HTML]{E1FFFF}91.50 & 29.00 & 34.50 & 35.50 & 14.00 & 20.70 & 67.00 & 29.50 & 43.70 & 26.50 & 12.00 & 11.00  & 56.75 & 25.92 & 32.03 \\
Llama3-70B & 80.00 & 58.00 & 66.10 & 23.00 & 7.00 & 6.50 & 66.00 & 49.50 & 56.20 & 30.50 & 9.50 & 14.70 & 62.00 & 27.00 & 39.30 & 45.50 & 18.50 & 16.90  & 51.17 & 28.25 & 33.28 \\
Llama3-8B & 70.50 & 45.00 & 55.30 & 17.00 & 5.50 & 5.20 & 64.00 & 44.50 & 53.10 & 8.50 & 1.00 & 2.20 & 37.50 & 13.00 & 21.50 & 37.00 & 10.50 & 9.00  & 39.08 & 19.92 & 24.38 \\
CodeLlama & 79.50 & 52.50 & 62.00 & 2.50 & 1.00 & 0.80 & 64.00 & 44.50 & 53.10 & 13.00 & 2.00 & 3.10 & 27.50 & 5.50 & 10.90 & 7.50 & 1.00 & 0.70 & 32.33 & 19.91 & 21.76 \\
\midrule
Average & \textbf{84.64} & \textbf{60.50} & \textbf{69.63} & 36.14 & 16.14 & 14.97 & \underline{74.29} & \underline{48.29} & \underline{55.64} & 36.21 & 13.50 & 19.57 & 62.36 & 28.07 & 40.46 & 46.50 & 16.64 & 15.41  & 55.09 & 32.19 & 36.82 \\
\bottomrule
\end{tabular}
\begin{tablenotes}
\definecolor{light cyan}{HTML}{E1FFFF}
\fboxsep1.5pt
\item[$\Phi$] \scriptsize Table cells with \colorbox{light cyan}{light cyan} background denote the highest performance in terms of CSR/CA/PR on each translation pair among various LLMs. Bold and underlined values represent the highest and second-highest average performance, respectively.
\end{tablenotes}
\end{threeparttable}
\vspace{-2em}
\end{table}

\subsection{RQ2: Translation Strategies}
\label{RQ2: Translation Strategies}
Our studied code translation strategy can be divided into three categories, i.e., direct, IR-based, and retrieval-based methods. As introduced in Section \ref{Studied Translation Strategies}, we propose three IR-based methods in this work, including \emph{IR(CoT)}, \emph{IR(pseudocode)}, and \emph{IR(summary)}. As for the retrieval-based method, a \emph{RA(method)} and \emph{RA(name)} are proposed for investigation, respectively. Figure \ref{rq2} presents the performance of various translation strategies in diverse LLMs on \textit{TransLibEval}. 
The key observations from our analysis are as follows:

\textbf{C++ as the Target PL:}
For C++-oriented translation, \emph{RA(name)} and \emph{IR(pseudocode)} strategies are the most effective across almost all evaluation metrics. \emph{RA(name)} excels due to C++'s strong convention of employing explicit, descriptive identifiers that precisely reveal function intent—a practice deeply ingrained in the C++ community for enhancing code readability and maintainability \cite{alsuhaibani2021naming, dos2006specifying}. As shown in Figure \ref{rq2-1}, through the \emph{RA(name)} strategy, our tool precisely retrieves effective posts from the knowledge base to produce the appropriate C++ implementation, exemplified by the \textit{CalculateDistance()} function. In our experiments, \emph{RA(name)} achieved a 41\% PR score, with Qwen performing particularly well in capturing these naming conventions.
\emph{IR(pseudocode)} effectively bridges the translation gap from library-rich PLs to library-limited ones.
Python and Java often rely on various off-the-shelf APIs to program, while C++ codes need to implement corresponding functionalities owing to the relatively limited libraries. Thus, LLMs tend to directly copy Python/Java APIs and libraries to implement the functionalities of the C++ side.
In this case, \emph{IR(pseudocode)} overcomes the above problem by isolating the influence of the original PLs but making LLMs focus exclusively on functionality logic.



\begin{boxK}
\small \faIcon{pencil-alt} \textbf{Finding 4:}
For C++-oriented translations, \emph{RA(name)} and \emph{IR(pseudocode)} strategies are the most effective. The intuitive method naming convention in the community facilitates retrieval with \emph{RA(name)} strategy, while \emph{IR(pseudocode)} isolates the API usage of original PLs, alleviating failures induced by library/API copy.
\end{boxK}

\begin{figure}[http]
\setlength{\abovecaptionskip}{0pt}
\centering
\includegraphics[width=\textwidth]{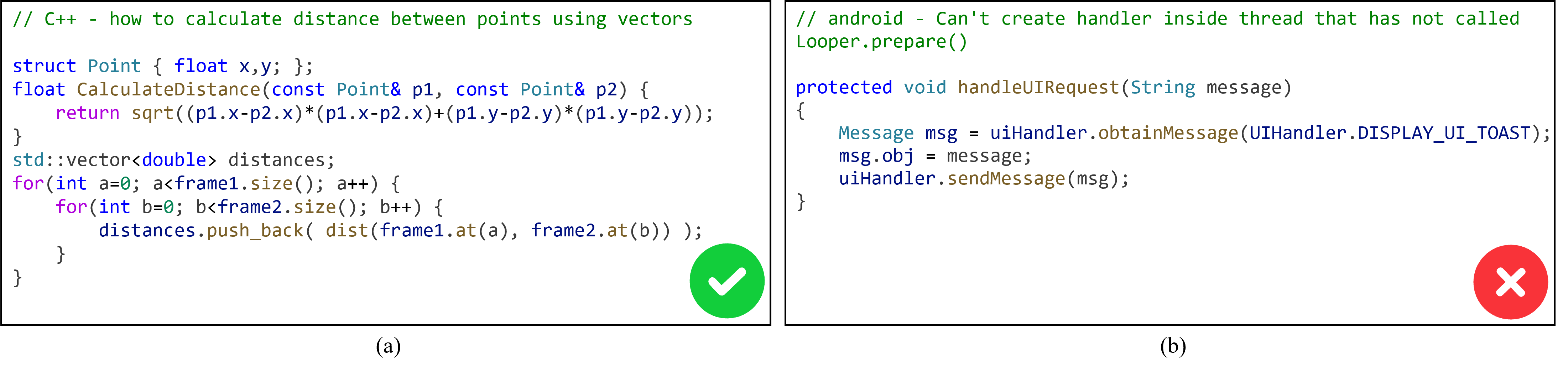}
\caption{Effectiveness of the \emph{RA(Name)} Strategy Across Different Target Languages.
(a) Successful retrieval of relevant C++ code patterns for \textit{CalculateDistance()}. (b) Unsuccessful retrieval in Java, returning irrelevant Android-specific code.}
\label{rq2-1}
\vspace{-0.5cm}
\end{figure}

\textbf{Java as the Target PL:}
Considering that the performance on CSR seems kind of complicated, as we cannot summarize which strategy performs best on Java-oriented translation, and CSR only evaluates on the compilation level, we focus on the evaluation metrics of CA and PR instead here. From this perspective, \emph{RA(method)} and \emph{Direct} strategies prove to be the most effective. \emph{RA(method)} can leverage rich contextual signals from method calls—such as parameter types, surrounding API sequences, and exception patterns—rather than relying solely on method names to retrieve reference implementations for LLMs. This is crucial in Java’s vast and varied ecosystem. For instance, when doing Python-to-Java translation, \emph{RA(name)} searches Java posts on Stack Overflow with a query of \textit{handle\_request}, a method name of a web-end Python program. However, an Android-end implementation for handling UI requests as shown in Figure \ref{rq2-1}(b) was obtained. This contradicts the Python program's intent, which is not for the mobile application, misleading LLMs' translation. 
As such, Retrieval based on full method context helps avoid such ambiguities and aligns well with Java’s heavy reliance on idiomatic API usage. On the other hand, the \textit{Direct} strategy remains effective, especially between Java and PLs with syntactic similarity such as C++. Shared structural conventions and relatively comparable standard libraries enable surprisingly accurate translations without additional retrieval.




\begin{boxK}
\small \faIcon{pencil-alt} \textbf{Finding 5:}
For Java-oriented translations, \emph{RA(method)} and \emph{Direct} strategies achieve the best results. The former allows for more contextual semantics to eliminate platform ambiguity, while the latter dominates in the C++ to Java translation owing to syntactic similarity. 
\end{boxK}

\textbf{Python as the Target PL:}
For Python-oriented translations, IR-guided strategies perform the best overall. A potential explanation is that the massive training base and language characteristics mentioned in Section \ref{RQ1: Overall Correctness} provide a relatively reliable generative ability for Python programs. Even without proactively injecting external knowledge, LLMs can induce numerous useful libraries for functionality implementation as shown in Table \ref{tab:lda_strategy_direction}, Section \ref{RQ3: Library-Dependency Awareness}. Among IR strategies, \emph{IR(pseudocode)} ranks the top-2 highest across almost all LLMs and evaluation metrics, because pseudocode as IR carries structured implementation logic for LLMs to follow, allowing LLMs to focus on core logic while abstracting away unnecessary syntactic complexity. 



\begin{boxK}
\small \faIcon{pencil-alt} \textbf{Finding 6:}
For Python-oriented translations, the IR strategy, especially \emph{IR(pseudocode)}, outperforms all others overall, owing to the advantages of LLMs themselves in Python programming and structured implementation logic offered by \emph{IR(pseudocode)}, making it the top choice for translating into Python.
\end{boxK}

\begin{figure}[htbp]
\vspace{-1.5em}
\setlength{\abovecaptionskip}{0pt}
\centering
\includegraphics[width=\textwidth]{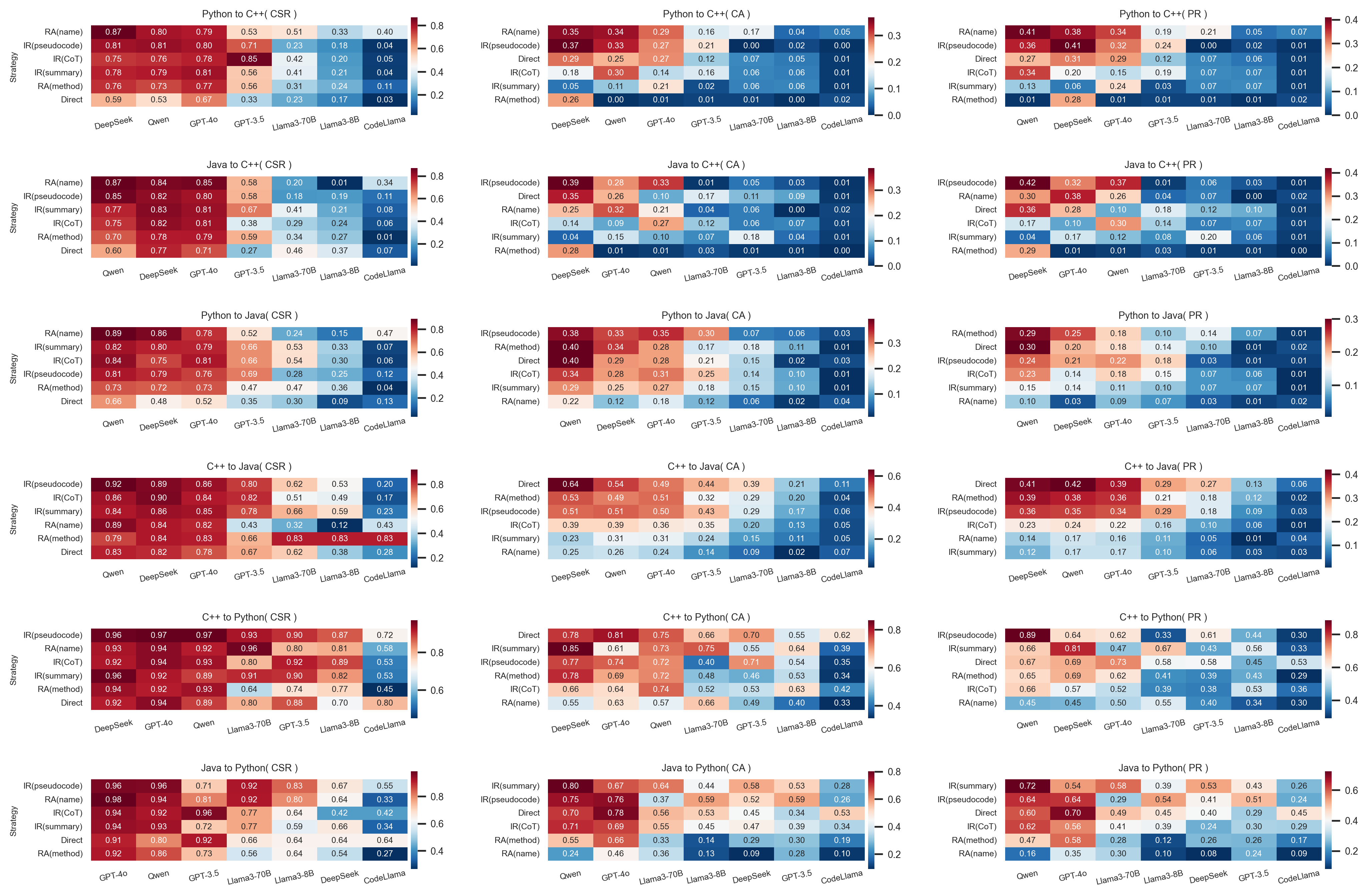}
\caption{Performance of Different Translation Strategies on \textit{TransLibEval}}
\label{rq2}
\vspace{-1.5em}
\end{figure}

\subsection{RQ3: Library-Dependency Awareness}
\label{RQ3: Library-Dependency Awareness}
Apart from the above correctness analysis, this section delves deeper into each translation from a library awareness perspective. 
Considering that the judgment of whether the LLMs identify useful libraries to accomplish the functionalities in the target PL side is hard to assess automatically, we opt to do so in a manual manner, which is the same as most previous related studies \cite{xue2024classeval-t,du2024evaluating}. 
Following their common workaround, we employed a stratified random sampling approach across all models, translation directions, and strategies. Samples were drawn proportionally from each stratum to form a balanced and unbiased sub-dataset for manual analysis. The data volume is 756, determined with a 95\% confidence level and a 5\% confidence interval, ensuring robust statistical coverage. To conduct a rigorous manual analysis, we determine that a translation satisfies LDA by examining 1) whether the translated program imports useful libraries, and 2) whether the APIs it invokes from imported libraries can implement the equivalent functionalities. The entire analysis process required a total of 70 person-hours. Four evaluators with 3–5 years of development experience carried out the manual analysis, with each sample being independently rated by two labelers. We employed Cohen’s kappa coefficient to evaluate inter-rater agreement, which is especially suitable for our task since the evaluation involves two distinct labels. The resulting kappa score of 0.71 indicates substantial agreement between the raters.
To simplify the analysis, Figure \ref{fig:rq3} displays the LDA results across different LLMs and translation directions. Considering the great correlation between translation strategies and translation directions during our manual analysis, we present Table \ref{tab:lda_strategy_direction} to simultaneously examine their TPL awareness performance.

\begin{figure}[htbp]
  \vspace{-0.5em}
  \centering
  \setlength{\abovecaptionskip}{0pt}
  \includegraphics[width=0.85\linewidth]{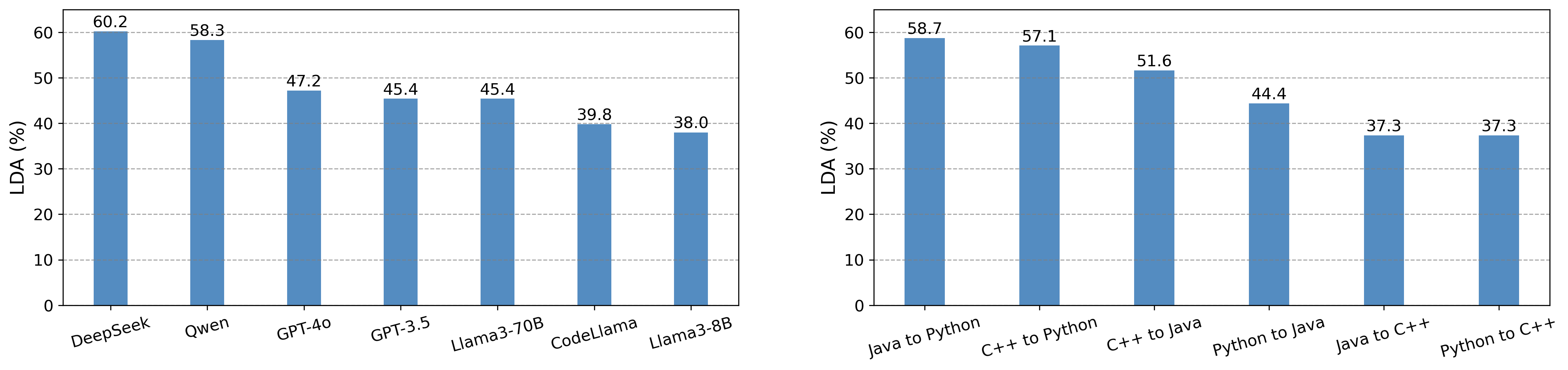}
  \caption{The Library-Dependency Awareness Analysis across Different LLMs and Translation Directions}
  \label{fig:rq3}
  \vspace{-1.0em}
\end{figure}

From the model perspective, commercial LLMs (e.g., DeepSeek, Qwen, GPT-4o, and GPT-3.5) consistently outperform other alternatives (e.g., Llama3-70/8B and CodeLlama) in identifying useful TPLs and invoking associated APIs by 28.51\% on average. In particular, DeepSeek leads with the highest LDA of 60.2\%. This trend can be attributed to their broader and more tremendous training bases, enormous parameter sizes, and advanced model architectures (e.g., MOE). 

Regarding translation directions, Python-oriented translations (i.e., Java/C++-to-Python) consistently demonstrate the highest LDA of 57.9\% on average, due to Python programs' massive training samples, broad availability of standard libraries, and natural language analogous features, lowering the threshold of LLMs to identify useful Python libraries and associated APIs.    
On the other hand, we also noticed that C++-oriented translations (i.e., Python/Java-to-C++) present more significant challenges for library dependency management, as evidenced by lower LDA scores of 37.3\% on average. A reasonable explanation is that standard libraries of C++ are limited to core language support, such as containers and algorithms. Implementing functionalities, such as AES symmetric encryption/decryption \cite{TransLib77:online}, requires collaboration of multiple TPLs or even partial implementation from scratch owing to the unavailability of certain libraries, which also significantly aggravates the difficulty of LLMs in identifying useful libraries.

In terms of strategies, LDA varies depending on different translation directions, especially the target PLs, as shown in Table \ref{tab:lda_strategy_direction}.
For Python as the target, IR-guided strategies, either \emph{IR(pseudocode)} or \emph{IR(summary)}, achieve the highest LDA scores, outperforming other counterparts by 31.60\%--37.93\%. When translating to C++, \emph{RA(name)} consistently keeps the best performance on library usage, surpassing counterparts by 35.14\% on the Python-to-C++ translation. As for Java-to-C++ translation, \emph{RA(name)} ties with \emph{IR(pseudocode)} but still excels other strategies by 75.98\%. Regarding the Java-oriented translation, \emph{Direct} dominates among various strategies, obtaining a significant lead by 37.26\%--66.68\%. The above conclusion of the LDA analysis among the different strategies is generally consistent with the correctness analysis across strategies in Section \ref{RQ2: Translation Strategies}, demonstrating in a more granular way the importance of identifying useful libraries/APIs in real-world code translation.   


\begin{boxK}
\small \faIcon{pencil-alt} \textbf{Finding 7:} Commercial LLMs outperform others in handling TPL dependencies. Python’s language features facilitate better dependency handling, while C++ presents greater challenges due to its complexity. Additionally, the choice of strategy significantly affects translation quality and dependency management.
\end{boxK}

\begin{table*}[t]
\centering
\caption{Strategy-wise LDA Rate (\%) across Translation Directions}
\vspace{-0.4cm}
\label{tab:lda_strategy_direction}
\tiny
\setlength\tabcolsep{4pt}
\begin{threeparttable}
\begin{tabular}{lcccccc}
\toprule
\multirow{2}{*}{\textbf{Strategy}} & \multicolumn{1}{c}{\textbf{C++ to Python}} & \multicolumn{1}{c}{\textbf{Python to C++}} & \multicolumn{1}{c}{\textbf{Java to Python}} & \multicolumn{1}{c}{\textbf{Python to Java}} & \multicolumn{1}{c}{\textbf{C++ to Java}} & \multicolumn{1}{c}{\textbf{Java to C++}} \\
\cmidrule(lr){2-7}
Direct & 57.14 & 38.10 & 71.43 & \cellcolor[HTML]{E1FFFF}66.67 & \cellcolor[HTML]{E1FFFF}66.67 & 38.10 \\
IR(CoT) & 66.67 & 33.33 & 47.62 & 42.86 & 57.14 & 23.81 \\
IR(pseudocode) & \cellcolor[HTML]{E1FFFF}71.43 & 33.33 & 57.14 & 38.10 & 47.62 & \cellcolor[HTML]{E1FFFF}52.38 \\
IR(summary) & 57.14 & 28.57 & \cellcolor[HTML]{E1FFFF}76.19 & 42.86 & 47.62 & 42.86 \\
RA(method) & 33.33 & 42.86 & 38.10 & 38.10 & 52.38 & 14.29 \\
RA(name) & 57.14 & \cellcolor[HTML]{E1FFFF}47.62 & 61.90 & 38.10 & 38.10 & \cellcolor[HTML]{E1FFFF}52.38 \\
\bottomrule
\end{tabular}

\begin{tablenotes}
\definecolor{lightcyan}{HTML}{E1FFFF}
\item \scriptsize Cells with \colorbox{lightcyan}{light cyan} background denote the highest LDA rate in each direction among all strategies.
\end{tablenotes}
\end{threeparttable}
\vspace{-0.6cm}
\end{table*}

\subsection{RQ4: Failed Cases Analysis}
\label{RQ4: Failed Cases Analysis}
This section meticulously examines failure cases that one of the SOTA LLMs (i.e., GPT-4o) made, aiming to identify persistent challenges in code translation involving various libraries, thereby highlighting areas where even the most advanced models require further refinement.

Our methodology here is a rigorous adaptation of the thematic analysis framework successfully employed by \textit{ClassEval-T} \cite{xue2024classeval-t}. For our evaluation, we dedicated 360 person-hours to thoroughly analyze 4,831 error cases selected from a single-time experiment with GPT-4o. We adopted the same structured, multi-stage process to ensure the accuracy and consistency of our findings. The process began with 3 experts with over five years of experience in Python, Java, and C++ who analyzed an initial sample set to create a comprehensive codebook for error categorization. This guide was then used by 7 experienced evaluators who assessed each case through a double-blind, dual-review process. In instances where evaluations diverged, the evaluators engaged in a structured discussion to reach a consensus, with unresolved disputes being escalated to the expert team for arbitration. To ensure the exhaustiveness of our findings, we conducted the categorization at a statement (e.g., API misuse)/token-level (e.g., indentation or symbol error) granularity, thereby a translation sample may consist of multiple types of errors.


Table \ref{tab:error_types} presents the evaluation results for 6 translation strategies on \textit{TransLibEval} across 6 translation directions. To better explain and clarify each type of error identified in our taxonomy, we have compiled a detailed 21-page Failed Cases Report, which is available in our repository \cite{TransLibEvalError2026}. From the overall distribution, \textbf{third-party reference errors (A)} account for the vast majority of issues, with an average proportion as high as 79.58\%, making them the primary cause of translation failure. In contrast, \textbf{syntactic errors (B)} and \textbf{code generation errors (C)} represent a smaller share, at 6.3\% and 11.1\% respectively. The proportions for \textbf{runtime errors (D)} and \textbf{other miscellaneous errors (E)} are very small, at just 2.48\% and 0.54\%. This clearly strengthens the position of \textit{TranLibEval} in evaluating library-centric code translation and exposing diverse library-related errors when LLM translates code.


Third-party reference errors primarily involve three aspects: \textbf{library-related errors (A-1)}, with an average proportion of 34.41\%; \textbf{API-related errors (A-2)}, with an average of 50.1\%; and \textbf{parameter-related errors (A-3)}, at 15.49\%. The high proportion of API-related errors indicates that API adaptation is the primary challenge in cross-PL translation, followed by library reference issues.
\textbf{Library-related errors (A-1)} stem from the improper mapping or handling of source-PL libraries in the target PL. The most common issue within this sub-category is \textbf{retaining the original PL's library (A-1-3)}, which accounts for 44.44\% of these errors. An example would be keeping \texttt{import numpy} in Java code. Furthermore, \textbf{calling an invalid library (A-1-1)} that doesn't exist or isn't supported in the target PL accounts for 36.44\%, while \textbf{missing required libraries (A-1-4)} account for 11.0\%, and \textbf{missing import statements (A-1-2)} for 8.12\%. 
\textbf{API-related errors (A-2)} are more challenging; the most prominent issue is \textbf{calling an irrelevant API (A-2-3)}, which makes up 55.31\% and often leads to a complete failure in code logic. Additionally, \textbf{retaining the original PL's API (A-2-1)} accounts for 33.24\%, while \textbf{missing an API (A-2-4)} and \textbf{calling a non-existent API (A-2-2)} account for 5.79\% and 5.66\%, respectively. 
\textbf{Parameter-related errors (A-3)} are mainly characterized by mismatches in parameter type, count, or return value when invoking APIs. Within this sub-category, \textbf{parameter type mismatch (A-3-1)} is the leading issue, accounting for 39.37\%. \textbf{Return value type errors (A-3-3)} represent 34.75\%, and \textbf{parameter count errors (A-3-2)} account for 25.89\%. 

The existence of syntactic errors indicates that LLMs still have shortcomings in syntax tree mapping between source and target PLs.
These errors primarily include three sub-categories: \textbf{Syntax structure errors (B-1)}, which account for 38.4\% and manifest as basic syntax issues like missing parentheses, improper indentation, or incomplete statements; \textbf{Language feature errors (B-2)}, which account for 21.86\% and stem from the failure to incorrectly translate unique syntax structures of the source PL (e.g., list comprehensions, pattern matching) into their equivalent expressions in the target PL; and \textbf{Object definition errors (B-3)}, which account for 39.73\%, referring to syntax errors occur in variable or function definition. 

Code generation errors, which account for 11.1\% of the total, are centered on the completeness and logical consistency of the output. Within this category, \textbf{the generation of non-code content (C-1)} accounts for 4.55\%, and the output mainly consists of explanatory text indicating the meaning of the source PL or text indicating that the task cannot be completed. In addition, \textbf{incomplete code generation (C-2)} and \textbf{contradictory code generation (C-3)} are the main issues, each accounting for approximately half of the cases. This suggests that the LLMs struggle to maintain code integrity at the generation level, especially with longer code or complex logic. \textbf{Runtime errors (D)}, which account for 2.48\%, are problems that are often difficult to expose during static analysis but appear during actual execution.  \textbf{Other errors (E)} account for only 0.54\%, mainly referring to miscellaneous issues that are difficult to classify and do not fall under major errors.

\begin{boxK}
\small \faIcon{pencil-alt} \textbf{Finding 8:} Third-party reference errors (A) are the dominant cause of code translation failures, accounting for a massive 79.58\% of all issues. In comparison, syntax (B) and code generation (C) errors are significantly smaller, while runtime (D) and other (E) errors are negligible. This indicates the effectiveness of \textit{TranLibEval} in the evaluation of TPL-involved code translation, exposing various kinds of failures that LLMs induce in the face of TPLs.  
\end{boxK}

\begin{table}[t]
\scriptsize
\setlength\tabcolsep{3pt}
\caption{Error categories vs. translation directions and strategies}
\vspace{-0.4cm}
\label{tab:error_types}
\begin{threeparttable}
\resizebox{\linewidth}{!}{%
\begin{tabular}{l>{\columncolor{red!32}}c>{\columncolor{red!22}}c>{\columncolor{red!12}}c>{\columncolor{red!12}}c>{\columncolor{red!12}}c>{\columncolor{red!12}}c>{\columncolor{red!22}}c>{\columncolor{red!12}}c>{\columncolor{red!12}}c>{\columncolor{red!12}}c>{\columncolor{red!12}}c>{\columncolor{red!22}}c>{\columncolor{red!12}}c>{\columncolor{red!12}}c>{\columncolor{red!12}}c>{\columncolor{yellow!32}}c>{\columncolor{yellow!12}}c>{\columncolor{yellow!12}}c>{\columncolor{yellow!12}}c>{\columncolor{blue!28}}c>{\columncolor{blue!8}}c>{\columncolor{blue!8}}c>{\columncolor{blue!8}}c>{\columncolor{green!12}}c>{\columncolor{brown!16}}c}
\toprule
\textbf{Categories} & \textbf{A} & \textbf{A-1} & \textbf{A-1-1} & \textbf{A-1-2} & \textbf{A-1-3} & \textbf{A-1-4} & \textbf{A-2} & \textbf{A-2-1} & \textbf{A-2-2} & \textbf{A-2-3} & \textbf{A-2-4} & \textbf{A-3} & \textbf{A-3-1} & \textbf{A-3-2} & \textbf{A-3-3} & \textbf{B} & \textbf{B-1} & \textbf{B-2} & \textbf{B-3} & \textbf{C} & \textbf{C-1} & \textbf{C-2} & \textbf{C-3} & \textbf{D} & \textbf{E} \\

\midrule
\multicolumn{26}{l}{\textbf{Direct}} \\ \midrule

\hspace{1.5em}C++ $\rightarrow$ Java & 73.31 & 19.26 & 6.76 & 1.35 & 8.78 & 2.36 & 32.77 & 10.81 & 0.68 & 17.91 & 3.38 & \textbf{21.28} & 9.12 & 5.07 & 7.09 & 3.04 & 1.69 & 1.35 & 0.0 & 14.19 & 0.34 & 10.14 & 3.72 & 6.76 & 2.7 \\
\hspace{1.5em}C++ $\rightarrow$ Python & \textbf{84.14} & 28.97 & 11.03 & 2.76 & 12.41 & 2.76 & 40.69 & 14.48 & 2.76 & 18.62 & 4.83 & 14.48 & 6.21 & 4.14 & 4.14 & 1.38 & 1.38 & 0.0 & 0.0 & 10.34 & 0.0 & 8.28 & 2.07 & 4.14 & 0.0 \\ 
\hspace{1.5em}Java $\rightarrow$ C++ & 72.57 & 27.88 & 10.84 & 2.21 & 10.84 & 3.98 & 29.87 & 12.39 & 0.22 & 15.04 & 2.21 & 14.82 & 6.42 & 3.32 & 5.09 & 3.54 & 0.88 & 2.21 & 0.44 & \textbf{14.6} & 0.0 & 13.72 & 0.88 & 8.19 & 1.11 \\ 
\hspace{1.5em}Java $\rightarrow$ Python & 69.7 & 24.24 & 7.58 & 1.52 & 11.36 & 3.79 & \textbf{45.45} & 12.88 & 0.76 & 31.82 & 0.0 & 0.0 & 0.0 & 0.0 & 0.0 & 1.52 & 1.52 & 0.0 & 0.0 & 12.12 & 0.0 & 6.82 & 5.3 & 13.64 & 3.03 \\ 
\hspace{1.5em}Python $\rightarrow$ C++ & 80.22 & 34.26 & 16.16 & 3.34 & 10.86 & 3.9 & 28.97 & 12.26 & 2.79 & 12.26 & 1.67 & 16.99 & 6.96 & 5.01 & 5.01 & 4.18 & 1.95 & 1.67 & 0.56 & 5.57 & 0.0 & 3.62 & 1.95 & 7.52 & 2.51 \\ 
\hspace{1.5em}Python $\rightarrow$ Java & 74.74 & \textbf{41.3} & 19.11 & 4.1 & 11.95 & 6.14 & 22.53 & 12.97 & 2.39 & 6.14 & 1.02 & 10.92 & 4.78 & 3.41 & 2.73 & \textbf{20.82} & 1.02 & 1.02 & 18.77 & 0.34 & 0.34 & 0.0 & 0.0 & 3.75 & 0.34 \\ 
\midrule
\multicolumn{26}{l}{\textbf{RA (method)}} \\ \midrule
\hspace{1.5em}C++ $\rightarrow$ Java & 70.3 & 30.3 & 11.52 & 2.42 & 11.52 & 4.85 & 35.15 & 13.94 & 3.64 & 16.36 & 1.21 & 4.85 & 0.61 & 0.61 & 3.64 & 1.21 & 1.21 & 0.0 & 0.0 & 15.15 & 0.0 & 15.15 & 0.0 & 12.73 & 0.61 \\ 
\hspace{1.5em}C++ $\rightarrow$ Python & 66.46 & 20.5 & 6.83 & 1.86 & 8.07 & 3.73 & 38.51 & 9.94 & 6.21 & 20.5 & 1.86 & 7.45 & 2.48 & 2.48 & 2.48 & 1.24 & 1.24 & 0.0 & 0.0 & 19.25 & 0.0 & 14.91 & 4.35 & 11.8 & 1.24 \\ 
\hspace{1.5em}Java $\rightarrow$ C++ & 68.94 & \textbf{32.95} & 17.05 & 3.41 & 9.47 & 3.03 & 22.35 & 10.98 & 3.79 & 0.0 & 7.58 & 13.64 & 4.55 & 3.41 & 5.68 & \textbf{18.56} & 1.14 & 12.5 & 4.92 & 12.5 & 0.0 & 0.38 & 12.12 & 0.0 & 0.0 \\ 
\hspace{1.5em}Java $\rightarrow$ Python & 73.44 & 17.19 & 4.69 & 1.56 & 8.59 & 2.34 & 34.38 & 9.38 & 7.03 & 17.97 & 0.0 & \textbf{21.88} & 11.72 & 1.56 & 8.59 & 1.56 & 1.56 & 0.0 & 0.0 & \textbf{21.09} & 0.78 & 17.19 & 3.12 & 2.34 & 1.56 \\ 
\hspace{1.5em}Python $\rightarrow$ C++ & 62.96 & 23.28 & 10.58 & 2.12 & 6.88 & 3.7 & 27.51 & 10.05 & 2.65 & 6.35 & 8.47 & 12.17 & 5.82 & 2.65 & 3.7 & 15.34 & 0.53 & 12.17 & 2.65 & 19.58 & 0.0 & 1.06 & 18.52 & 2.12 & 0.0 \\ 
\hspace{1.5em}Python $\rightarrow$ Java & \textbf{77.14} & 27.43 & 9.14 & 2.29 & 13.71 & 2.29 & \textbf{45.71} & 15.43 & 1.71 & 25.14 & 3.43 & 4.0 & 0.57 & 2.86 & 0.57 & 3.43 & 0.57 & 2.29 & 0.57 & 18.29 & 0.0 & 5.71 & 12.57 & 0.57 & 0.57 \\ 
\midrule
\multicolumn{26}{l}{\textbf{RA (name)}} \\ \midrule
\hspace{1.5em}C++ $\rightarrow$ Java & 84.38 & 28.82 & 10.07 & 2.08 & 13.89 & 2.78 & \textbf{50.69} & 14.93 & 0.69 & 34.38 & 0.69 & 4.86 & 1.74 & 1.74 & 1.39 & 3.12 & 1.74 & 0.35 & 1.04 & 7.64 & 0.0 & 4.86 & 2.78 & 4.51 & 0.35 \\ 
\hspace{1.5em}C++ $\rightarrow$ Python & \textbf{89.89} & 24.47 & 7.45 & 1.6 & 13.3 & 2.13 & 48.94 & 14.89 & 0.53 & 30.85 & 2.66 & 16.49 & 6.91 & 3.72 & 5.85 & 0.0 & 0.0 & 0.0 & 0.0 & 8.51 & 0.0 & 3.19 & 5.32 & 1.06 & 0.53 \\ 
\hspace{1.5em}Java $\rightarrow$ C++ & 87.39 & 34.78 & 13.48 & 3.04 & 13.04 & 5.22 & 34.35 & 13.91 & 2.61 & 16.52 & 1.3 & \textbf{18.26} & 8.7 & 3.48 & 6.09 & \textbf{6.52} & 1.74 & 0.0 & 4.78 & 3.48 & 0.0 & 1.3 & 2.17 & 1.74 & 0.87 \\ 
\hspace{1.5em}Java $\rightarrow$ Python & 75.3 & 16.27 & 3.01 & 0.6 & 11.45 & 1.2 & 48.8 & 11.45 & 1.2 & 36.14 & 0.0 & 10.24 & 6.02 & 1.2 & 3.01 & 3.61 & 3.61 & 0.0 & 0.0 & \textbf{20.48} & 0.0 & 19.28 & 1.2 & 0.0 & 0.6 \\ 
\hspace{1.5em}Python $\rightarrow$ C++ & 86.9 & \textbf{36.24} & 13.97 & 3.06 & 13.97 & 5.24 & 38.43 & 14.85 & 0.87 & 21.4 & 1.31 & 12.23 & 4.8 & 3.06 & 4.37 & 5.68 & 2.18 & 0.0 & 3.49 & 6.11 & 0.0 & 2.62 & 3.49 & 0.87 & 0.44 \\ 
\hspace{1.5em}Python $\rightarrow$ Java & 86.72 & 29.46 & 10.79 & 2.49 & 13.28 & 2.9 & 50.21 & 14.11 & 2.9 & 32.78 & 0.41 & 7.05 & 2.9 & 1.66 & 2.49 & 2.9 & 0.83 & 0.0 & 2.07 & 6.64 & 0.0 & 4.15 & 2.49 & 3.32 & 0.41 \\ 
\midrule
\multicolumn{26}{l}{\textbf{IR (CoT)}} \\ \midrule
\hspace{1.5em}C++ $\rightarrow$ Java & 89.28 & 31.37 & 12.33 & 2.68 & 13.67 & 2.68 & 39.14 & 14.21 & 0.8 & 24.13 & 0.0 & 18.77 & 5.9 & 5.9 & 6.97 & 3.49 & 2.14 & 0.27 & 1.07 & 4.02 & 1.61 & 0.0 & 2.41 & 2.95 & 0.27 \\ 
\hspace{1.5em}C++ $\rightarrow$ Python & 76.34 & 16.13 & 3.23 & 1.08 & 10.75 & 1.08 & 48.92 & 10.75 & 3.76 & 34.41 & 0.0 & 11.29 & 3.76 & 3.76 & 3.76 & 0.54 & 0.0 & 0.0 & 0.54 & \textbf{22.58} & 0.0 & 3.23 & 19.35 & 0.0 & 0.54 \\ 
\hspace{1.5em}Java $\rightarrow$ C++ & 61.11 & 18.22 & 8.0 & 1.78 & 6.44 & 2.0 & 18.67 & 7.33 & 2.0 & 7.11 & 2.22 & 24.22 & 9.78 & 6.22 & 8.22 & \textbf{28.89} & 12.67 & 1.78 & 14.44 & 8.67 & 0.0 & 0.0 & 8.67 & 0.44 & 0.89 \\ 
\hspace{1.5em}Java $\rightarrow$ Python & 75.52 & 9.79 & 1.4 & 0.7 & 6.29 & 1.4 & 33.57 & 8.39 & 4.2 & 16.78 & 4.2 & \textbf{32.17} & 14.69 & 6.29 & 11.19 & 0.0 & 0.0 & 0.0 & 0.0 & 20.98 & 0.0 & 0.0 & 20.98 & 3.5 & 0.0 \\ 
\hspace{1.5em}Python $\rightarrow$ C++ & 68.86 & 29.71 & 13.14 & 2.86 & 11.14 & 2.57 & 25.43 & 11.71 & 0.57 & 11.71 & 1.43 & 13.71 & 4.86 & 4.29 & 4.57 & 23.43 & 11.71 & 2.29 & 9.43 & 6.29 & 0.0 & 0.0 & 6.29 & 0.86 & 0.57 \\ 
\hspace{1.5em}Python $\rightarrow$ Java & \textbf{94.51} & \textbf{35.71} & 12.09 & 2.75 & 17.03 & 3.85 & \textbf{51.1} & 18.13 & 0.55 & 31.87 & 0.55 & 7.69 & 2.75 & 2.2 & 2.75 & 2.75 & 0.0 & 0.0 & 2.75 & 1.65 & 0.0 & 1.65 & 0.0 & 0.55 & 0.55 \\ 
\midrule
\multicolumn{26}{l}{\textbf{IR (pseudocode)}} \\ \midrule
\hspace{1.5em}C++ $\rightarrow$ Java & 83.24 & 21.79 & 5.59 & 1.12 & 13.41 & 1.68 & \textbf{54.75} & 15.08 & 0.0 & 36.87 & 2.79 & 6.7 & 1.68 & 1.68 & 3.35 & 2.79 & 0.0 & 0.0 & 2.79 & 13.41 & 0.0 & 7.82 & 5.59 & 0.56 & 0.0 \\ 
\hspace{1.5em}C++ $\rightarrow$ Python & 70.83 & 21.53 & 8.33 & 2.08 & 9.72 & 1.39 & 34.03 & 9.72 & 4.17 & 20.14 & 0.0 & 15.28 & 5.56 & 5.56 & 4.17 & 0.0 & 0.0 & 0.0 & 0.0 & \textbf{29.17} & 0.0 & 7.64 & 21.53 & 0.0 & 0.0 \\ 
\hspace{1.5em}Java $\rightarrow$ C++ & 85.06 & 38.31 & 15.58 & 3.25 & 14.61 & 4.87 & 36.69 & 15.26 & 0.32 & 20.45 & 0.65 & 10.06 & 3.57 & 3.57 & 2.92 & 5.52 & 0.65 & 1.3 & 3.57 & 6.49 & 0.32 & 3.25 & 2.92 & 1.62 & 1.3 \\ 
\hspace{1.5em}Java $\rightarrow$ Python & 72.03 & 24.58 & 9.32 & 2.12 & 9.75 & 3.39 & 24.58 & 10.17 & 2.97 & 11.02 & 0.42 & \textbf{22.88} & 11.02 & 3.81 & 8.05 & \textbf{13.56} & 2.12 & 11.44 & 0.0 & 13.98 & 0.0 & 9.75 & 4.24 & 0.0 & 0.42 \\ 
\hspace{1.5em}Python $\rightarrow$ C++ & \textbf{90.44} & \textbf{39.34} & 19.85 & 4.04 & 11.76 & 3.68 & 31.25 & 12.5 & 1.1 & 15.44 & 2.21 & 19.85 & 9.19 & 3.68 & 6.99 & 1.47 & 0.37 & 0.37 & 0.74 & 5.88 & 0.37 & 4.41 & 1.1 & 1.1 & 1.1 \\ 
\hspace{1.5em}Python $\rightarrow$ Java & 89.18 & 31.6 & 12.12 & 2.6 & 13.85 & 3.03 & 48.92 & 14.72 & 0.43 & 32.9 & 0.87 & 8.66 & 3.03 & 2.16 & 3.46 & 1.73 & 0.87 & 0.43 & 0.43 & 8.23 & 1.73 & 5.63 & 0.87 & 0.43 & 0.43 \\ 
\midrule
\multicolumn{26}{l}{\textbf{IR (summary)}} \\ \midrule
\hspace{1.5em}C++ $\rightarrow$ Java & \textbf{98.44} & 24.92 & 6.23 & 1.25 & 15.89 & 1.56 & \textbf{68.22} & 15.89 & 3.12 & 49.22 & 0.0 & 5.3 & 2.49 & 1.25 & 1.56 & 1.25 & 0.0 & 0.31 & 0.93 & 0.31 & 0.0 & 0.31 & 0.0 & 0.0 & 0.0 \\ 
\hspace{1.5em}C++ $\rightarrow$ Python & 95.98 & \textbf{36.61} & 12.05 & 2.68 & 18.3 & 3.57 & 54.91 & 18.75 & 2.68 & 33.48 & 0.0 & 4.46 & 1.79 & 1.34 & 1.34 & 0.45 & 0.0 & 0.45 & 0.0 & 1.79 & 0.0 & 1.79 & 0.0 & 1.34 & 0.45 \\ 
\hspace{1.5em}Java $\rightarrow$ C++ & 77.65 & 28.24 & 12.94 & 2.65 & 10.0 & 2.65 & 36.47 & 12.35 & 1.18 & 15.59 & 7.35 & \textbf{12.94} & 5.59 & 2.65 & 4.71 & 10.29 & 7.94 & 0.29 & 2.06 & 11.47 & 0.0 & 1.47 & 10.0 & 0.29 & 0.29 \\ 
\hspace{1.5em}Java $\rightarrow$ Python & 87.18 & 17.52 & 2.56 & 0.85 & 13.25 & 0.85 & 60.68 & 13.68 & 4.27 & 42.74 & 0.0 & 8.97 & 3.42 & 2.56 & 2.99 & 0.0 & 0.0 & 0.0 & 0.0 & \textbf{11.54} & 0.0 & 0.43 & 11.11 & 1.28 & 0.0 \\ 
\hspace{1.5em}Python $\rightarrow$ C++ & 71.97 & 26.23 & 10.54 & 2.24 & 10.76 & 2.69 & 33.18 & 11.66 & 0.45 & 18.61 & 2.47 & 12.56 & 4.71 & 3.59 & 4.26 & \textbf{15.47} & 10.09 & 5.38 & 0.0 & 8.74 & 0.22 & 0.0 & 8.52 & 2.47 & 1.35 \\ 
\hspace{1.5em}Python $\rightarrow$ Java & 88.54 & 31.85 & 12.74 & 2.55 & 13.38 & 3.18 & 54.78 & 15.92 & 4.46 & 24.2 & 10.19 & 1.91 & 0.64 & 0.64 & 0.64 & 8.28 & 3.18 & 2.55 & 2.55 & 3.18 & 1.91 & 1.27 & 0.0 & 0.0 & 0.0 \\ 
\bottomrule
\end{tabular}%
}
\end{threeparttable}
\end{table}

Figure \ref{rq4-1} illustrates a horizontal comparison of error analysis among different translation strategies, where the Third-Party Reference (TPR) errors consistently account for the largest proportion. In terms of overall error volume, the \emph{Direct} strategy and the \emph{IR(CoT)} strategy yielded significantly higher counts, ranging from 1,600 to over 1,700 cases. In contrast, retrieval-augmented methods remained in a lower range, between 1,000 and 1,400 errors.

In the \emph{Direct} strategy, TPR errors account for 74.77\% of all errors, with a significant portion arising from library-related errors, indicating that the \emph{Direct} strategy fails in the library importation stage, not to mention calling APIs, which leads to the top tier in terms of the number of errors.
The \emph{IR(summary)} strategy shows the highest rate of API-related errors (A-2), particularly the use of calling an irrelevant API (A-2-3), accounting for 30.63\% of its TPR errors, suggesting that while summarization aids in capturing high-level logic, it still struggles with accurate API alignment across programming environments. 
The \emph{IR(CoT)} strategy exhibits a higher proportion of parameter-related errors (A-3), especially parameter type mismatches (A-3-1) and return value errors (A-3-3), comprising 21.00\% of its TPR issues. The incorporation of chain-of-thought reasoning appears to introduce additional complexity in parameter mapping, despite its detailed reasoning process. 

Among all strategies, retrieval-augmented strategies, especially \emph{RA(method)}, produce the fewest errors, benefiting from broader contextual retrieval for better API and library matching.
Although \emph{IR(pseudocode)} yields the best results among IR strategies, it only performs on par with \emph{RA(name)}. The former often loses critical implementation details during abstraction, while the latter is limited by retrieval inaccuracies.
These observations underscore the persistent difficulty of aligning third-party references in cross-lingual code translation and highlight the specific improvement directions and precautions for use of the above translation strategies. 

\begin{figure}[http]
\setlength{\abovecaptionskip}{0pt}
\centering
\hspace*{-0.2cm}
\includegraphics[width=1\textwidth]{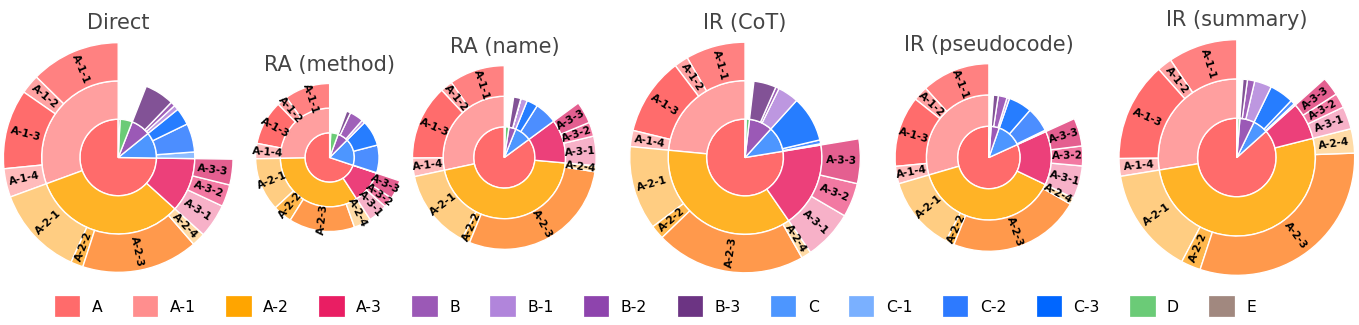}
\caption{A Horizontal Comparison of Error Analysis among Different Translation Strategies}
\label{rq4-1}
\end{figure}

\begin{boxK}
\small \faIcon{pencil-alt} \textbf{Finding 9:}  
Although third-party reference errors are prevalent across all translation strategies, their distributions reveal distinct patterns: the \emph{Direct} strategy suffers most from library errors (A-1), \emph{IR(summary)} peaks in API errors (A-2), and \textit{IR(CoT)} exhibits high parameter errors (A-3). Retrieval-augmented methods, particularly \emph{RA(method)}, achieve the lowest error rates. While \emph{IR(pseudocode)} performs best among IR variants, it only matches \emph{RA(name)}, with limitations in abstraction and retrieval scope, respectively. These findings underscore the need for more robust handling of external library references and precautions for use of the above translation strategies.



\end{boxK}


\vspace{-0.8em}
\section{Implications}


\textbf{Implications for researchers:}
(1) As the inaugural study to concentrate on TPL utilization in code translation, this work brings to light fundamental limitations in existing benchmarks, which typically lack high categorical coverage and large-scale TPL-targeted code samples. (2) According to the well-known failures involving TPLs/APIs \cite{pan2024lost,yang2024exploring,xue2024classeval-t}, we introduce five translation strategies under two widely applicable and highly motivated methodological systems, i.e., IR-guided and retrieval-augmented, and make comparisons with the \emph{Direct} strategy in RQ2. Among the five strategies, \emph{IR(pseudocode)}, \emph{RA(method)}, and \emph{RA(name)} are the first time to be applied in the code translation field for addressing TPL-related failures, offering basic design principles and usage experience for this area's future research. (3) Afterwards, we explain the results of RQ1-2 with multiple views (LLMs, translation directions, and strategies) by delving into the library dependency awareness, significantly improving academia's understanding of why certain LLMs perform better with certain strategies on certain translation directions.
(4) Furthermore, RQ4 identifies a spectrum of failure patterns covering 19 categories, where 11 of them are unique to library-centric scenarios, such as calling an invalid library (A-1-1) and calling an irrelevant API (A-2-3). These issues are frequently obscured in prior studies \cite{yang2024exploring,xue2024classeval-t,Wang2025ApiRAT,luo2025integrating,wang2024repotransbench,ibrahimzada2024repository} due to their benchmarks' limitations mentioned above in (1). Thus, our work provides more explicit directions for future code translation research.

\textbf{Implications for practitioners:}
(1) We constructed the first TPL-targeted code translation benchmark, \textit{TransLibEval}, which consists of 200 real-world TPL-involved samples for industrial evaluation. (2) Through \textit{TransLibEval}, we conducted extensive experiments to investigate the performance of diverse LLMs with different translation strategies on various translation directions among mainstream PLs, i.e., Python, Java, and C++, in RQ1-3, revealing the heterogeneous advantages of diverse LLMs and translation strategies as well as offering suggestions and explanations for developers' practical usage.      
(3) Moreover, 19 error types are manually categorized from 4,831 failures of GPT-4o, precisely reminding practitioners
to notice its potential risks when deploying for TPL-intensive scenarios. Besides, the follow-up strategy-wise analysis further offers insightful suggestions from the perspectives of their frequent errors, helping practitioners to understand each strategy's limitations beyond their advantages in RQ2. 

\vspace{-0.8em}
\section{Threats to Validity}

\textbf{Threats to external validity} relate to the generality of the findings and the failure taxonomy~\cite{liu2024exploring,xue2024automated}. Our evaluation is limited to code translation between Python, Java, and C++, excluding other PLs. However, \textit{TransLibEval} is composed of larger-scale, richer-category coverage, practical TPL-centric tasks than previous work (e.g.,~\cite{khan2024xcodeeval,xue2024classeval-t,ibrahimzada2024repository}), making our results more applicable to real-world development. The dominance of Python, Java, and C++ in industry and academia ~\cite{TIOBE2025} also confirms the practical implications of our findings. To construct the failure taxonomy, we analyzed a corpus of 4,831 distinct error cases from one of the SOTA LLMs. This extensive analysis, spanning all three PLs, six strategies, and six translation directions, guided by thematic analysis, substantially minimizes threats to the generality of our conclusions.

\textbf{Threats to internal validity} mainly involve potential data leakage problems and manual handling procedures of this work \cite{xue2024automated,ibrahimzada2024repository}.  
To mitigate the data leakage problem, we manually constructed \textit{TranLibEval}, which significantly reduces the possibility of being seen by LLMs. Moreover, a recent study \cite{yang2025rethinking} proved with extensive experiments that code translation tasks are almost unaffected by data leakage owing to their unpaired features of source-target codes in LLMs' pre-training. 
\textit{TransLibEval} is intended to incorporate a self-refreshing design that regularly updates via automated program mutation, thereby maintaining its relevance, rigor, and data leakage-free over time.
For tasks involving manual construction and inspection, multiple expert annotators performed a double-blind review following thematic analysis principles to minimize subjectivity and error. Hence, the above threats are minimal.

\textbf{Threats to construct validity} concern the comprehensiveness of evaluation metrics \cite{Roy2023Survey, Coker2023Methodology}. 
To achieve this, we exert PR to evaluate the test case pass rate and CA to assess the pass rate of the whole test suite.
Besides, we also examine their CSR from a compilation perspective. Moreover, we leverage LDA to investigate the ability of LLMs to correctly identify useful TPLs when translating code. We do not include literal consistency, e.g., exact match accuracy, for assessment because code translation involving third-party dependencies is way more challenging, and no sample can be translated exactly the same with ground truths, making such evaluation meaningless. In addition, semantically equivalent programs are not necessarily consistent in literal expressions. Hence, considering the above diverse evaluation workarounds, this threat is limited.

\vspace{-0.3em}
\section{Conclusion and Future Work}
This work constructs the first benchmark for evaluating TPL-targeted code translation, namely \textit{TransLibEval}, and introduces 6 highly motivated translation strategies for assessment. Extensive experiments on 7 recent LLMs of diverse categories and sizes demonstrate their weaknesses on TPL-involved code translation and uncover the heterogeneous advantages of each strategy, offering valuable guidance for practical usage.   
Finally, we proceeded with a thorough manual analysis and categorization of LLMs' translation failures on third-party dependency samples, as the first attempt, shedding light for researchers and practitioners to facilitate their studies and applications.

Based on our findings, our future work aims to enhance the performance of LLMs in TPL-targeted code translation by further optimizing retrieval or IR-guided strategies.
Additionally, we will extend \textit{TransLibEval} to a broader range of PL pairs with strong industrial relevance, such as C++ to Rust.

\section{Acknownledgement}
This work was partially supported by the National Natural Science Foundation of China (Grant Nos. 62502283 and U24B20149), the Natural Science Foundation of Shandong Province (Grant No. ZR2024QF093),  and the Young Talent of Lifting Engineering for Science and Technology in Shandong, China (Grant No. SDAST2025QTB031).

\noindent\textbf{Data Availability:} We open-sourced the replication package and \textit{TransLibEval} benchmark at \cite{TransLib77:online}.


\newpage
\bibliographystyle{ACM-Reference-Format}
\bibliography{sample-base}
\end{document}